%% file: main.tex
\documentclass[a4paper,fleqn,10pt]{article}
\usepackage[T1]{fontenc}
\usepackage{lmodern} 
\PassOptionsToPackage{normalem}{ulem}
\input{macros/macros}

\usepackage{todonotes}
\usepackage{subcaption}
\usepackage{xcolor}
\usepackage{tikz}

\usepackage[style=numeric,sorting=none,citestyle=numeric-comp]{biblatex}
\usepackage{hyperref}
\hypersetup{pdfborder={0 0 0},unicode=true}

\preprint{ 
    FERMILAB-PUB-26-0091-T \\
    IPPP/26/018 \\ 
    MCNET-26-02 
}    

\title{Robust Calibration of Non-Perturbative\\[9pt] Models with History Matching}

\renewcommand{\thefootnote}{\fnsymbol{footnote}}

\author{
    Andrew Iskauskas$^1$\footnote[1]{andrew.iskauskas@durham.ac.uk},
    Max Knobbe$^2$\footnote[2]{mknobbe@fnal.gov}, 
    Frank~Krauss$^3$\footnote[3]{frank.krauss@durham.ac.uk},
    Steffen Schumann$^4$\footnote[4]{steffen.schumann@phys.uni-goettingen.de}
}

\institute{
    $^1$ Department of Mathematical Sciences, Durham University, Durham DH1 3LE, UK\\[3pt]
    $^2$ Fermi National Accelerator Laboratory, Batavia, IL 60510, USA\\[3pt]
    $^3$ Institute for Particle Physics Phenomenology, Durham University, Durham DH1 3LE, UK\\
    $^4$ Institut f\"ur Theoretische Physik, Georg-August-Universit\"at G\"ottingen, 37077 G\"ottingen, Germany
}
\hypersetup{
  pdfauthor={
    Andrew Iskauskas, Max Knobbe, Frank Krauss, Steffen Schumann
    },
  pdftitle={
    Sherpa Calibration
    },
  pdfkeywords={
    QCD, Event Generators, Non-Perturbative Physics, Tuning, Calibration
    }
}
\date{February 2026}

\newcommand{\Exp}{\mathbb{E}}
\newcommand{\Var}{\text{Var}}
\newcommand{\Cov}{\text{Cov}}
\newcommand{\vect}[1]{\mathbf{#1}}

\begin{document}

\maketitle

\begin{abstract}
\noindent We apply, for the first time, Bayes Linear Emulation and History Matching to the calibration of non-perturbative models in Monte Carlo event generators.
In contrast to the usual approach of ``Monte Carlo tuning'', History Matching does not result in best-fit plus ellipsoidal parameter uncertainty estimates but instead identifies all parameter space regions that are consistent with data.
This approach leads to a systematic and robust quantification of parametric uncertainties in the models, especially in those challenging cases where different, possibly disjoint, regions of parameter space deliver similar results, which are usually not properly treated with current methodology.
We highlight the power of this method with the hadronisation models available through \Sherpa: the built-in cluster fragmentation \Ahadic and string fragmentation through an interface to \Pythia.  
\end{abstract}

\newpage
\tableofcontents
\newpage

\renewcommand{\thefootnote}{\arabic{footnote}}

\section{Introduction}

The Monte Carlo event generators (MCEGs) \Herwig~\cite{Bellm:2019zci,Bellm:2025pcw}, \Pythia~\cite{Bierlich:2022pfr}, and \Sherpa~\cite{Sherpa:2019gpd,Sherpa:2024mfk} are indispensable tools for experimental analyses at particle colliders such as the \LHC.
Aiming at complete descriptions of individual particle collisions, they combine first-principles perturbative calculations with phenomenological models of the non-perturbative phases of the events~\cite{Buckley:2011ms,Campbell:2022qmc}. 

The latter of these poses a challenge, since such models result in relatively high-dimensional parameter spaces which have to be calibrated to data.
This is a computationally very demanding if not prohibitive task, and it is of course not unique to particle physics.
The usual response to this obstacle is to construct a fast approximation to the parameter dependence of the model predictions, a surrogate~\cite{Buckley:2009bj,Krishnamoorthy:2021nwv,LaCagnina:2023yvi}; in the \textsc{Professor} framework~\cite{Buckley:2009bj}, widely used in particle physics, this is realised through polynomials in the parameters for each data bin used in the calibration.  
The aim then is to find parameter combinations that describe data sufficiently well. However, currently employed standard techniques usually do not reliably find different local minima simultaneously.
As a consequence, the stopping criteria used to move away from the minimum often need to be manually adjusted to satisfy intuition.
Once a point in parameter space that minimises loss or maximises some goodness-of-fit measure is found, local variations around this ``optimal set'' -- usually an ellipsoid if $\chi^2$ quality measures are used -- are employed to evaluate parametric uncertainties.
This method can be refined, for example, by repeating the procedure multiple times, with different selections of data or applying different relative weights to them in the ``tuning'' process~\cite{Knobbe:2023njd}~\footnote{
  Ref.~\cite{LaCagnina:2023yvi} considered the construction of the full posterior distribution of the data given the (surrogate) generator model, which allows the efficient exploration of the parameter space and the extraction of statistically interpretable parameter-uncertainty estimates.}. 
In general, the notion of a ``singular best fit'' might obscure the fact that quite distinct parameter combinations can produce simulations of comparable quality; especially when extrapolating the calibration results to new dynamic domains, this may result in overly optimistic uncertainty estimates.

This problem is convincingly addressed by emulation-based History Matching (HM), a method that is well adapted to find all local minima and to rigorously identify parameter sets of comparable quality, suitable for uncertainty estimation. 
HM has been used for a variety of different settings where complex computer models and their parameters need to be calibrated to real-world, experimental data, including in the oil industry \cite{craig1997pressure}, climate science \cite{wilson2022varying}, galaxy formation \cite{vernon2014galaxy}, and atomic physics \cite{vernon2022nuclear}. 
One of the most prominent uses of HM has been in the field of epidemiology, where computationally expensive simulators~\cite{andrianakis2017efficient, scarponi2021tbhiv, iskauskas2024HPV, krauss2022june} lead to a high-dimensional output space, where both model and observation are subject to uncertainties in how they represent physical reality, and where simulator outputs are not deterministic.

In this work, we detail the first ever application of Bayes Linear emulation and History Matching to the non-perturbative models in MCEGs in particle physics.
This represents not only a new way for the particle physics community to obtain parameter calibrations for their simulations, but circumvents the problems of scalability that arise from the high resolution of both models and data in particle physics.
As poster child of non-perturbative models, in this paper we focus on hadronisation, which describes the transition of quarks and gluons, the quanta of perturbative quantum chromodynamics (QCD), to primary hadrons, 
which subsequently decay. The corresponding models are based on a qualitative understanding of the dynamics of this transition, and usually rely on around $20$ parameters to describe the relatively intricate and multi-faceted data, which typically stems from high-energy electron--positron annihilations into hadrons.
In particular, we will calibrate the hadronisation models available in \Sherpa, namely the cluster fragmentation model of \textsc{Ahadic}~\cite{Chahal:2022rid} and the string fragmentation model available through an interface to \textsc{Pythia 8}~\cite{Bierlich:2022pfr}. 
Keeping all other aspects of the simulation (the hard-process component and the parton-shower model) identical, allows us to not only robustly estimate parametric uncertainties of the two models, but simultaneously quantify the uncertainty related to the choice of physics model for hadronisation. 

In Section~\ref{sec:hme} we review the salient fundamentals of Bayes Linear emulation and History Matching. 
In Section~\ref{sec:SHERPA} we detail how they apply to our specific application, describe the \Sherpa simulation setups, the data used for calibration, and discuss the performance of the emulator. 
With these concepts in hand, Section~\ref{sec:results} demonstrates the results obtained from emulation and HM with \Sherpa and \Sherpa\!\!+\Pythia, including considerations of the differences between the two models' outputs and the structure of the final non-implausible space. 
Finally, in Section~\ref{sec:conclude} we discuss limitations of and possible extensions to this work.

\section{Emulation and History Matching}\label{sec:hme}
The hadronisation models in \Sherpa, and any other MCEG, require a process of determining acceptable parameter combinations which allow them to reflect the physical reality (a process often referred to as \emph{tuning}). 
The fine detail of both model and data render this process extremely difficult both conceptually and, despite the simulation efficiency, computationally, because the parameter space of the models is high-dimensional. 
The space of interest here comprises around $20$ parameters for both the \Ahadic and \Pythia hadronisation. 
The high-dimensional nature of the space poses problems for many established methods of calibration: to explore even just the corners of the space would require over half a million simulator runs with hundreds of thousands of simulated events per run - a computationally prohibitive task. 
Therefore, exploration of the full space is infeasible without recourse to some efficient method of calibration, which ideally requires a minimum of the costly simulator runs.

Optimisation approaches~\cite{akiba2019optuna} may allow a comparatively small number of simulator runs to find moderately acceptable parameter combinations, but require the determination of an appropriate cost function to minimise -- this in itself may be hard to define when the number of outputs under consideration is large, as it is in this application. 
Furthermore, optimisation in high input dimension is prone to settling in local minima, particularly on the boundaries of the space~\cite{brynjarsdottir2014learning}, and it is difficult to truly explore the \emph{full} parameter space of matches to data, even with repeated random restarts of the optimisation process. 
More sophisticated approaches such as MCMC- or ABC-based calibration~\cite{mckinley2018approximate} can provide a more rigorous calibration result, providing a posterior distribution on the parameter space, but these methods depend on access to large numbers of simulator runs.
They also suffer heavily from the `curse of dimensionality'~\cite{fernandez2020curse}, and will always return some posterior even in those cases where fundamental conflicts between the model and observation indicate that no acceptable match to data exists. 

Instead, HM~\cite{craig1997pressure} operates on the principle of removing the unacceptable parts of parameter space rather than targeting the `good' parts. 
The uncertainty structure of HM allows us to replace simulator runs with predictions from an emulator: a statistical surrogate for simulator output~\cite{bower2010galaxy, andrianakis2015bayesian, williamson2013history}.

\subsection{The History Matching Framework}

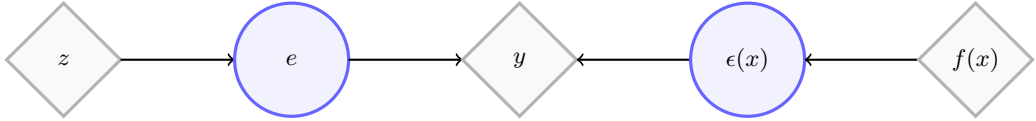
\begin{figure}
\centering
\begin{tikzpicture}[scale = 1.5]
\filldraw[color = gray!60, fill = gray!5, very thick] (0,0) -- (0.5, 0.5) -- (1, 0) -- (0.5, -0.5) -- cycle;
\node at (0.5, 0) {\small $f(x)$};
\draw [thick, ->] (1,0) -- (2,0);
\filldraw[color = blue!60, fill = blue!5, very thick] (2.5,0) circle(0.5);
\node at (2.5, 0) {\small $\epsilon(x)$};
\draw [thick, ->] (3, 0) -- (4, 0);
\filldraw[color = gray!60, fill = gray!5, very thick] (4,0) -- (4.5, 0.5) -- (5, 0) -- (4.5, -0.5) -- cycle;
\node at (4.5, 0) {\small $y$};
\draw [thick, ->] (5, 0) -- (6, 0);
\filldraw[color = blue!60, fill = blue!5, very thick] (6.5, 0) circle(0.5);
\node at (6.5, 0) {\small $e$};
\draw [thick, ->] (7,0) -- (8, 0);
\filldraw[color = gray!60, fill = gray!5, very thick] (8, 0) -- (8.5, 0.5) -- (9, 0) -- (8.5, -0.5) -- cycle;
\node at (8.5, 0) {\small $z$};
\end{tikzpicture}
\caption{The structure linking the real-world process $y$ to the observations $z$ and the simulator $f(x)$ via the observational error $e$ and the model discrepancy $\epsilon(x)$.}
\label{fig:tikzeg}
\end{figure}

To motivate the philosophy behind HM and emulation, we first outline a schematic that connects the real-world phenomena to our observational data, shown in Figure~\ref{fig:tikzeg}.

Suppose we have a real process $y$ of interest -- in this work this is the underlying physics which manifests itself in the particle collisions -- and we want to connect it to a physical reality. 
For observations $z$ (the experimental results) of this physical reality, we link the two via a schematic relation
\begin{equation*}
    z = y + e\,,
\end{equation*}
where $e$ denotes some random quantity encoding the discrepancy between observation and reality, related to the impact of  measurement errors or any systematic experimental bias.
Simply put, observations represent but do not perfectly reflect physical realities.
We also have to accept that computer simulators of real-world processes are imperfect representations of reality. 
Let the simulator output at a given parameter combination $x$ in our parameter space $\mathcal{X}$ be denoted $f(x)$. 
Then
\begin{equation*}
    y = f(x) + \epsilon(x),
\end{equation*}
where $\epsilon(x)$ is a (possibly parameter-dependent) random quantity which we term the \emph{model discrepancy} and which reflects the imperfection of the simulator. 
The link between our simulator and our observations is therefore
\begin{equation*}
    z = f(x) + e + \epsilon(x).
\end{equation*}

This structure, while conceptually simple, highlights the two key sources of uncertainty in our calibration task. 
The choice of model discrepancy $\epsilon(x)$ can be motivated by domain knowledge, elicitation from a collection of pilot runs, or other physical considerations: discussions of the structure of model discrepancy, for example, are detailed in~\cite{goldstein2013assessing}~\footnote{
    There are further arguments for introducing this disconnect between simulator output and reality, particularly when we consider the fundamental meaning behind the parameters in the simulator, but we do not linger on this philosophical question here.}.
A successful calibration targets those $x$ where $f(x)$ and $z$ are in alignment, \emph{up to} the fundamental discrepancies between simulation and observation, by answering the following question.

\noindent\emph{Given observed data corresponding to a simulator output, what combinations of input parameters could give rise to output consistent with this observation within the stated uncertainties.}

HM aims to construct the answer via a simple restatement: given a collection of simulator runs and observational data, what combinations of input parameters can we confidently rule out even having accounted for the inherent uncertainties? 
To make this approach concrete, we introduce an \emph{implausibility measure} $I(x)$ on the parameter space
\begin{equation}\label{eq:nativeIMP}
    I(x)^2 = \frac{\Exp[f(x)-Z]^2}{\Var[f(x)-Z]}.
\end{equation}
A high value of $I(x)$ suggests that the difference between simulator output and observational data is so large that, even accounting for uncertainty, it is very unlikely that $x$ is a suitable candidate parameter set: we term such a point \emph{implausible}. 
Conversely, a low implausibility does not guarantee that a parameter combination is `good', in the traditional sense: $I(x)$ can be low either because the difference between $f(x)$ and $z$ is small, or because the combined uncertainties $\Var[f(x)-Z]$ are large. 
These points are termed \emph{non-implausible} or \emph{not-yet-ruled-out}; the choice of language here is intentional and underlines that a low implausibility is not the same as, for instance, a minimum of a cost function or maximisation of a likelihood function and says nothing about the `goodness-of-fit' of the point under question.

It is important to distinguish the implausibility measure and other measures used in calibration -- particularly Gaussian likelihoods and reduced $\chi^2$ statistics, despite the similarity in their forms.
In our construction of implausibility, we make no distributional assumptions about the model or the observation-generating process, nor do we assume a particular form for the model itself. The non-implausible region is imbued with no concept of coverage, in the statistical sense; unlike the $\chi^2$ measure, for example, we do not have recourse to standard arguments for what constitutes a `high' value of $I(x)$.
However, we may follow Pukelsheim's rule~\cite{pukelsheim1994three}, which states that for \emph{any} unimodal distribution $95\%$ of the probability mass lies within $3\sigma$ of the mean, and choose $I=3$ as the threshold for high implausibility.
We revisit the differences between the implausibility measure and the reduced $\chi^2$ statistic in Section~\ref{sec:results}, and a more general discussion may be found in \cite{vernon2010rejoinder}.

The HM framework gives a clear avenue for carefully specifying our error structure and incorporating it into a rigorous search of the parameter space. 
Clearly, if we are able to remove all implausible regions of parameter space, then, by complementarity, we are left with the full space of parameter combinations that could have given rise to our observations. 
While Eq.~\eqref{eq:nativeIMP} is completely general, we are at liberty to state any reasonable assumptions we have about the structure of the system in question. 
Here, we make the following statements:
\begin{itemize}
    \item The simulator is deterministic\footnote{The simulator used here is deterministic in the sense that $f(x)$ is fixed in the infinite-statistics limit. In practice, each run of a Monte Carlo event generator is an average over finitely many points, introducing the usual Monte Carlo expression. We treat this stochastic variability as an additional contribution to the model discrepancy.}: the quantity $f(x)$ does not change given fixed input $x$, so is not a random quantity. We therefore have $\Exp[f(x)-Z]=f(x)-\Exp[Z]$;
    \item Our observational error is not expected to contain systematic bias, so that $\Exp[e]=0$ and hence $\Exp[Z]=z$;
    \item Model discrepancy and observational error are independent, and so
    \begin{equation*}
    \Var[f(x)-Z] = \Var[z - e - \epsilon(x) - Z] = \Var[e]+\Var[\epsilon(x)].
    \end{equation*}
\end{itemize}

\subsection{Linking Emulators to History Matching}
While the HM philosophy allows us to incorporate our beliefs and provides a means for finding all parameter contributions that match to data, it still depends on simulator runs $f(x)$. 
We are, however, at liberty to replace our simulator with an appropriate surrogate with lower computational expense. 
Provided we understand the disconnect between these two objects, this can be easily incorporated into the implausibility measure. 
Such a surrogate is referred to as an \emph{emulator}~\cite{craig2001bayesian}. 
Emulators have been applied in a variety of contexts and using a number of different structures, for instance, the family of Gaussian Process emulators \cite{conti2009gaussian, andrianakis2012effect} or, in the context of high-energy particle physics, for example, as surrogates for scattering amplitudes~\cite{Maitre:2021uaa,Janssen:2023ahv,Maitre:2023dqz,Bahl:2024gyt} or for traditional calibration methods~\cite{Buckley:2009bj,Ilten:2016csi,Krishnamoorthy:2021nwv,AlKadhim:2025npf}.

In this work, we use Bayes Linear emulators~\cite{goldstein2007bayes} which have low burden of specification. 
An emulator $g(x)$, emulating a simulator output $f(x)$, takes the form
\begin{equation}\label{eq:emeq}
g(x) = \sum_{i=1}^p \beta_i h_i(x) + u(x),
\end{equation}
where the $h_i(x)$ are $p$ known basis functions in the parameters $x$ with $\beta_i$ their coefficients. 
They account for the global response of the output $g(x)$ across parameter space $\mathcal{X}$, and $u(x)$ is a smooth structure that accounts for local variability around this global response. 
While the functions $h_i(x)$ are assumed known, the coefficients $\beta_i$ and stationary process $u(x)$ are random variables that require prior specification~\footnote{
    Despite being `known', in the statistical sense, we still need to choose the structure of the $h_i(x)$. 
    This is model- and application-dependent: we discuss the choices for this particular application in Section~\ref{sec:SHERPA}.}.

In high-dimensional settings, we often modify this structural form via the introduction of \emph{active variables} $x_A$ for an output $f(x)$.
The rationale behind the modification is that, for a given output, it is unlikely that every parameter contributes substantively to the output response, and so the basis functions $h_i(x)$ and stationary process $u(x)$ do not need to depend on the variables which have little to no influence.
In this case, Eq.~\eqref{eq:emeq} becomes
\begin{equation*}
    g(x) = \sum_{i=1}^p \beta_i h_i(x_A) + u(x_A) + w(x),
\end{equation*}
where $w(x)$ is some residual term (often referred to as a `nugget') that accounts for the small residual variability due to the exclusion of the inactive parameters from the other terms.
This can be seen as a form of dimensional reduction in the parameter space: since we are likely to obtain different active variable subsets for each output, we might use the determination of active variables as an indication of output clustering, or (in the extreme case where a variable is inactive for all outputs) a motivation for classifying a varying parameter as uninformative for the calibration task at hand~\footnote{
    It is worth noting that the active variable subset for a given output will change over waves of HM; the parameters that govern large-scale behaviour over a large initial non-implausible region are unlikely to be the same as those that influence small-scale changes once the scope of the region of interest is reduced.}.
The use of active variables is eminently sensible in this approach, since we have a moderately high input dimension and very high output dimension, and the effect of considering active variables is something we will return to in Section~\ref{sec:SHERPA}.

The Bayes Linear (BL) approach asks only that we decide on first- and second-order specifications for these quantities (i.e., expectations and variances). 
Therefore, we need to specify:
\begin{itemize}
    \item A $p$~dimensional vector $\Exp[\beta]$ of expectations and a $p\times p$ matrix $\Var[\beta]$ of corresponding (co)variances;
    \item The expectation and variance $\Exp[u(x)]$ and $\Var[u(x)]$ for any parameter combination $x$. 
    We will decompose the specifications for $u(x)$ further. 
    We will assume it to be zero-mean, and assume a covariance structure $\Cov[u(x),u(x^\prime)]=\sigma^2 c(x, x^\prime)$, with $c(\cdot,\cdot)$ a suitably chosen correlation function;
    \item The covariance $\Cov[\beta, u(x)]$ at all points $x$.
    \item Where we consider active variables, we must also specify the form of $w(x)$: often we assume this is zero-mean, uncorrelated with the other terms in the emulator, and with variance $\Var[w(x)]=\sigma_w^2$.
\end{itemize}

We pause here for a discussion on notation.
It is important to note that, in the Bayes Linear paradigm, expectation is considered `primitive' and statements about it can be made without recourse to any underlying distribution, either implicitly or explicitly.
One can demonstrate that this definition of expectation accords exactly in properties with the more familiar interpretation, and second-order quantities $\Var[\cdot]$ and $\Cov[\cdot, \cdot]$ follow naturally.
A full discussion of the philosophy behind Bayes Linear statistics is beyond the scope of this work, but we refer an interested reader to \cite{goldstein2007bayes} for a fuller treatment.

One advantage of the BL approach is that these specifications are interpretable and can be easily obtained via expert elicitation or analysis of a small collection of simulator runs. 
We may think of $\Exp[\beta]$ and $\Var[u(x)]=\Cov[u(x),u(x)]$ as our prediction of the simulator output and our corresponding uncertainty, while the term $\Cov[u(x),u(x^\prime)]$ encodes the level of connectedness we expect our model output to have across differing parameter combinations. $\Var[\beta]$ can be naturally thought of as our uncertainty in the accuracy of our prior specifications about the regression coefficients, with off-diagonal terms encoding the extent to which we think misspecification in one coefficient may affect our certainty about the others, and $\Cov[\beta, u(x)]$ similarly relates our beliefs about the regression surface to that of the weakly stationary process~\footnote{In practice, provided we have sufficiently many model runs we may treat the coefficients as `known', so that $\Var[\beta]=0$. In the absence of strong beliefs about how we expect the global (regression) to correlate with the local variability, we often specify $\Cov[\beta, u(x)]=0$, though the adjusted covariance $\Cov_D[\beta, u(x)]$ does not remain zero.}.
We also expect the covariance to be higher when $\vert x-x^\prime\vert$ is small, and attenuate as the distance between them increases. 
This leads us to the use of an isotropic correlation function $c(x,x^\prime)=c(\vert x-x^\prime|)$, of which many interpretable examples exist~\cite{rasmussen2003gaussian}~\footnote{
    Often, the choice of a particular isotropic correlation function is a reflection of any beliefs about smoothness of the output. 
    If we believe it to be infinitely differentiable, then an exponential-squared function is appropriate while the Matern class of correlation functions allows us to specify a fixed level of differentiability (the Matern $n+1/2$ correlation function is exactly $n$-times differentiable).}.

Having determined prior specifications for such an emulator, we wish to expose it to data and obtain informed predictions across the full parameter space. 
This bears similarities to the general concept of Bayesian updates, with the caveat that we have not provided distributional specifications for our random quantities and therefore may not apply Bayes' Theorem in the normal fashion~\footnote{
    If we were convinced that our second-order specifications correspond exactly to those that would arise from the appropriate multivariate normal, then the standard Bayesian update follows those calculations for a Gaussian process emulator~\cite{vernon2022bayesianjune}. 
    However, for complex systems we must consider carefully whether we truly believe such a distributional specification is accurate, or even appropriate.}. 
We can instead proceed by using the BL update formulae. 
Suppose we have data $D$, arising from a collection of simulator runs. Then we may consider the updated expectation and variance of the emulator at any parameter combination $x^*$, denoted $\Exp_D[g(x^*)]$ and $\Var_D[g(x^*)]$~\footnote{In analogue with discussion around the prior specifications, these expectations and variances are not computed with respect to an underlying data distribution; the subscript $D$ indicates the data update computed via these Bayes Linear update equations.}, which can be calculated as
\begin{eqnarray}
    \Exp_D[g(x^*)] &=& \Exp[g(x^*)] + \Cov[g(x^*), D]\Var[D]^{-1}(D-\Exp[D]), \label{eq:ED}\\
    \Var_D[g(x^*)] &=& \Var[g(x^*)]-\Cov[g(x^*),D]\Var[D]^{-1}\Cov[D,g(x^*)].\label{eq:VarD}
\end{eqnarray}
Note that these update equations are simply matrix equations: for a fixed collection of simulator runs $D$, the computationally expensive matrix inversion is a single off-line calculation.
The emulator allows us to make predictions of the simulator output across the parameter space quickly~\footnote{
    For most applications, emulator predictions have typical computational times of the order $10^{-7}$ seconds, outperforming all but the fastest simulators.  
    They are also easily parallelisable.} 
using a relatively small collection of actual simulator runs. 
While the expression $\mathbb{E}_D[g(x^*)]$ represents predictions of the simulator output at $x^*$, as opposed to the simulator output at $x^*$ itself, the emulators also automatically provide the corresponding uncertainty in their prediction via $\Var_D[g(x^*)]$.

Using the emulator rather than the simulator modifies the HM process detailed above and induces a change in the implausibility measure:
\begin{equation}\label{eq:impproper}
I(x)^2=\frac{(\Exp_D[g(x)]-z)^2}{\Var_D[g(x)]+\Var[e]+\Var[\epsilon(x)]}.
\end{equation}
We see that our interpretation of implausibility now includes also any emulator uncertainty we might have in our predictions. 
While we can explore the parameter space faster by orders of magnitude via recourse to the emulators, this additional source of uncertainty means that, without modification to our process, the interpretation of the non-implausible region no longer simply depends on discrepancies induced by reality or the modelling task.

To solve this problem in our approach, we apply multiple iterations (termed \emph{waves}) during HM. 
Suppose that we train an emulator to an initial collection of simulator runs across some space $\mathcal{X}$. 
The regions of parameter space deemed implausible by this emulator will be implausible even with a less uncertain emulator; we therefore take a sample of points from the non-implausible region, $\tilde{\mathcal{X}}$, and use these to obtain a further collection of simulator runs. 
A second-wave emulator trained on this new ensemble of runs will have lower uncertainty, and, as a consequence will be able to rule out further parts of the space $\tilde{\mathcal{X}}$ as implausible, resulting in a new non-implausible space $\hat{\mathcal{X}}$. 
We may repeat this procedure, typically until the emulator uncertainty is subdominant to the other, fundamental uncertainties in the system: the final non-implausible space therefore represents those regions of parameter space that remain acceptable to the most accurate possible emulators of the simulator on that space.

Before explicitly outlining the iterative HM and emulation algorithm, we must briefly discuss one aspect of matching simulators to data. 
We frequently have multiple observations and multiple corresponding outputs from our simulator that we wish to match to simultaneously; we must therefore emulate this collection of outputs appropriately. 
Suppose that we have $N$ outputs $f_1(x), \dots, f_N(x)$ with corresponding observations $z_1,\dots, z_N$. 
Then the most straightforward approach is to emulate each output individually, creating emulators $g_1(x),\dots, g_N(x)$; as a consequence, we have a number of different choices for the structure of an implausibility measure. 
If we have strong beliefs about the correlation between given outputs, we may define a multivariate implausibility measure given the vector of observations $\vect{z}=(z_1, \dots, z_N)^\top$ and emulators $\vect{g}(x)=(g_1(x),\dots, g_N(x))^\top$
\begin{equation}\label{eq:multivarimp}
I(x)^2 = (\Exp_D[\vect{g}(x)]-\vect{z})\Var_D[\vect{g}(x)-\vect{z}]^{-1}(\Exp_D[\vect{g}(x)]-\vect{z}),
\end{equation}
akin to a Mahalanobis distance on the space. 
In the absence of strong beliefs about correlations, we may instead define some composite measure on univariate implausibilities, such as maximum implausibility
\begin{equation*}
I_M(x) = \max_{i=1,\dots,N} I_i(x)
\end{equation*}
from which second-maximum, third-maximum, \dots, $N$th-maximum (i.e. minimum) measures follow accordingly. 

Even with these considerations, we may still be concerned about the computational cost when the number of outputs is large: while the emulator evaluations are extremely efficient and the process of training the emulators is equally computationally light, for very high-dimensional output spaces the process of emulation can lose the computational advantage over a simulator which automatically returns all outputs of interest. 
However, we note one aspect of the implausibility measure touched upon in a different context above: if we are able to rule out a putative parameter combination as a result of high implausibility with respect to one output, then the inclusion of other outputs will not reverse or refute this exclusion. 
Hence, at each wave of the process we are at liberty to include only those outputs which are most informative for the process of ruling out parameter space, including those more subtly dependent on changing parameter values at a later stage and remaining confident in the robustness of our history match. 
This is in stark contrast to methods such as optimisation, where the focus on finding `good' parameter combinations requires us to include all sources of information throughout the process. 
We leverage this advantage in Section~\ref{sec:results} for the hadronisation models available in \Sherpa, but for concreteness we detail the algorithm behind waves of emulation and HM below.

\begin{enumerate}
\item Let $\mathcal{X}_0$ be the initial domain of interest and set $k=1$;
\item Generate an appropriate design for a set of runs over the non-implausible space $\mathcal{X}_{k-1}$;
\item Identify the collection of informative outputs, $Q_k$, chosen via domain expert knowledge, principle variable analysis \cite{cumming2007dimension}, or a combination of both, and obtain the corresponding simulator evaluations by running the simulator for this design of points;
\item Construct new, more accurate, emulators defined only over $\mathcal{X}_{k-1}$ for the collection $Q_k$;
\item Use the emulators to calculate implausibility across the entirety of $\mathcal{X}_{k-1}$, discarding points with high implausibility to define a smaller non-implausible region $\mathcal{X}_k$;
\item If any of our predetermined stopping criteria have been met, continue to Step 7. Otherwise, repeat the algorithm from Step 2 for wave $k+1$;
\item Generate a large number of acceptable runs from $\mathcal{X}_k$, sampled appropriately.
\end{enumerate}

We conclude this section with one final comment, which pertains to Step 6 above. 
There is no single stopping criterion for the HM process, and it depends on the simulator, observations, and non-implausible parameter space obtained. 
We postpone the discussion of stopping conditions considered in this application to Section~\ref{sec:SHERPA}, save to highlight one general possibility: one reason for terminating a history match is if $\mathcal{X}_k=\emptyset$ for some wave $k$. 
Such a result would demonstrate a fundamental disconnect between the output of our simulator and observational reality, suggesting that the model as currently specified is incapable of providing acceptable matches to data. 
That this outcome is a natural consequence of the approach demonstrates another key difference between traditional methods of calibration: in particular, methods that return a posterior will always return the `least bad' parameter subset regardless of whether `least bad' and `good' are synonymous in the given application~\cite{vernon2010rejoinder}.
While one can, and should, apply post-hoc analysis and diagnostics on the resulting posterior to identify such possible fundamental mismatches to data or model specification,
this is not necessary in HM -- if the model cannot match to the observational data, we will be made aware of that and can perform directed investigations guided by the implausibility statements to find where the incompatibilities between simulator and observation occur, providing an additional means for model diagnostics.

The above section highlights the main differences between HM and other methods for calibration of complex models: the inherent uncertainty structure makes it amenable to the introduction of predictions from statistical surrogates in place of simulator evaluations. 
It also allows for an intuitive and robust approach to selectively and variously including observations of interest at each wave, allowing us to focus on those aspects of the modelling problem that matter at each stage. 
Furthermore, the Bayes Linear emulators have an intuitive, interpretable structure that can be specified easily with recourse to expert judgement, and allow a computationally efficient exploration of the (high-dimensional) parameter space. 
We will now apply these concepts to the hadronisation models available in the \Sherpa event-generator framework.

\section{Application to Hadronisation Modelling}\label{sec:SHERPA}
In this section we detail the application of HM and emulation to the case of hadronisation model calibration. 
To this end, we compile information on the considered parameter spaces and simulator setups, and the data used to match to.
We also discuss details of the used emulators and measures to judge their quality. 

\subsection{Parameter space, Observations, and Emulation}
\Sherpa~\cite{Sherpa:2019gpd,Sherpa:2024mfk,Gleisberg:2008ta,Gleisberg:2003xi} is a multi-purpose event generator designed for the simulation of collision events at high-energy particle colliders. 
Its modular structure allows \Sherpa to mix and match different, physically equivalent components in order to systematically assess variations introduced by alternative modelling approaches. 
Separate modules are used to describe different physics regimes.
Simulations typically start with the hard collision, described in perturbative QCD, continue with the QCD evolution, usually via parton showers, and then transition to the non-perturbative regime. 
In this work, we focus on the last step of this pipeline, and in particular on the modelling of the non-perturbative parton-to-hadron transition region. 
More specifically, we study the formation of primary hadrons and the corresponding hadronisation modelling.
Within the \Sherpa framework, there are two separate options to model this transition: the built-in cluster-hadronisation model \Ahadic~\cite{Chahal:2022rid} and an interface to the string-fragmentation model implemented in \Pythia 8~\cite{Bierlich:2022pfr}. 
While cluster models are based on the concept of colour preconfinement~\cite{Amati:1979fg}, the central construction paradigm of string fragmentation is the linear potential of QCD between quarks~\cite{Andersson:1983ia} (for a more detailed review, see for example~\cite{Buckley:2011ms,Campbell:2022qmc}). 
The decays of unstable hadrons are handled by the decay module of \Sherpa for the case of \Ahadic and by \Pythia for the string fragmentation. 
Both models can be invoked on the same perturbative simulation input, which allows us to derive an unbiased estimate for the uncertainty related to the choice of the hadronisation model. 
However, the perturbative accuracy of the hard-process calculation and the parton-shower simulation, as well as design choices in the latter, have a residual impact on the hadronisation process; see, for example, Ref.~\cite{Hoang:2024nqi}. 
As a consequence, a hadronisation model needs to be re-calibrated when used in conjunction with a different event generator, as is the case for the Lund-string fragmentation when used in combination with \Sherpa. 

Hadronisation-model parameters can best be constrained by comparison with precision data from measurements of hadronic final-state observables in electron--positron annihilation, in particular data from experiments at the Large Electron Positron collider (LEP) at $\sqrt{s}=91.2$\,\UGeV. 
For the corresponding event simulations, we use a pre-release version of \Sherpa 3.1, and calculate the hard process $e^+e^-\rightarrow q\bar{q}+\{0,1\}\mathrm{jets}$ at next-to-leading order (NLO) accuracy, using tree-level matrix elements from its built-in generators \textsc{Amegic}~\cite{Krauss:2001iv} and \textsc{Comix}~\cite{Gleisberg:2008fv}, together with one-loop amplitudes from \textsc{OpenLoops2}~\cite{Buccioni:2019sur} using the \textsc{Collier} library~\cite{Denner:2016kdg}. 
The matrix elements are matched to \Sherpa's default parton shower~\cite{Schumann:2007mg} using the internal \MEPSatNLO formalism~\cite{Hoeche:2012yf}. 
The renormalisation scale entering the calculation is set accordingly~\cite{Hoeche:2009rj}. 
For the jet-separation criterion in the merging procedure we use $\mathrm{log}_{10}(Q^2_{\mathrm{cut}}/s)=-2$~\cite{Hoeche:2009rj}. 
We employ two-loop running for the strong coupling $\alpha_s$ with an input value of $\alpha_s(m^2_Z) = 0.118$.

The data to which the hadronisation models are matched to consist of measurements from various LEP experiments, obtained via \textsc{Rivet}~\cite{Bierlich:2024vqo}. 
Each of these experiments provides multiple observables, e.g.\ event-shape distributions, fragmentation functions, identified-particle momentum spectra, and hadron multiplicities.  
In total, we identified $432$ outputs -- each output being a single bin of one of the observed histograms -- for which data were reliable enough to warrant inclusion in the calibration problem. 
For a detailed list of observables, see Appendix~\ref{app:observable_selection}. 
The observations have well-defined measurement errors, allowing a straightforward specification of our observational uncertainty $e$.
For \Ahadic we consider $19$ parameters to calibrate against the data, while for \Pythia we selected $23$ that specify the Lund-string model. 
See Tables~\ref{tab:params_ahadic} and \ref{tab:params_pythia} in Appendix~\ref{app:parameter_selection} for the respective lists of parameters chosen. 
For each model an initial collection of around $1000$ runs, covering the parameter space, provided a starting point for investigation.  
These initial waves were created using a coarse space-filling exploration of the relevant parameter space, serving to verify that the model behaved sensibly -- that is, that it produced results in a reasonable amount of time -- and provide the emulators with the largest possible scope of model behaviours to emulate.
Since each simulator run of \Sherpa results from an ensemble of Monte Carlo simulations of finite statistics, \Sherpa is not a deterministic model in the strictest sense. 
However, pilot runs and subsequent calibration runs demonstrated that the stochastic variability is relatively stable across the parameter space: we therefore chose to include the variability as an $x$-independent contribution to the model discrepancy, re-evaluated at each wave of the HM process~\footnote{One could include the stochastic variability as a varying quantity in its own right; for a description of the principles behind variance emulation, see for example \cite{iskauskas2024HPV}.}. 
It is important to note that while this variability is taken to be $x$-independent, it is still a significant contributor to the overall uncertainty for some of the observations. In just over $8\%$ of observations, this variability is larger than the observational uncertainty, and in the most extreme cases it is five times larger. It is therefore crucial that it be taken into account, lest we incorrectly rule out some regions of parameter space. We will see the effect of this variability in Section~\ref{sec:results}.
Furthermore, to reflect the fact that no model is a perfect replication of reality, we included an additional source of model discrepancy based on the magnitude of the observations, varying between $1\%$ and $3\%$ of the magnitude of the observation, dependent on the output in question and the progression through the HM waves. 
The choice of the magnitude of this additional term, referred to as `external' model discrepancy, is difficult to derive from exploratory model analysis. 
As it encodes the extent to which we believe our model to be an imperfect representation of reality, we are unlikely to be able to directly quantify the level of imperfection; however, one may use the results of diagnostic checks and consideration of the non-implausible region generated to assure oneself that the choice is appropriate.
For more details on model discrepancy, and in particular possible means by which order-of-magnitude statements about external model discrepancy may be made, we refer the reader to \cite{goldstein2013assessing}.

As mentioned in Section~\ref{sec:hme}, we need not incorporate each observation in every wave; indeed, examination of the outputs for each analysis suggests a level of smoothness and consistency between outputs which makes emulating all outputs unnecessary.
We instead choose to demand that each of the $29$ individual experimental observables is represented at each wave, but that only a per-observable subset of bins (observations) is chosen, based on principal variable analysis~\cite{cumming2007dimension} and requiring that at least $95\%$ of the output variability of each observable be explained, in analogue with the setting of a threshold on variance explained in Principal Component Analysis. 
These choices were validated and reviewed at each wave by consideration of the simulator outputs for all observations, providing a diagnostic check on our assumption of smoothness. 
This process identified between $50$ and $70$ outputs at each wave, and the results indicated the expectation of smoothness of the experimental data. 
Over the complete history match of \Ahadic, $127$ of the complete set of outputs were emulated; the more involved calibration of \Pythia emulated $185$ outputs in total. 
Note that while these particular subsets of outputs were used to train the emulators, we evaluate the quality of the eventual fits with respect to the full $432$ observational quantities available to us. 
One might see this as a useful diagnostic on our beliefs about the model structure and our choices of observations to emulate, as matching to the remaining $304$ and $247$ observations, respectively, is by no means guaranteed. 
Interestingly, the outputs emulated in the \Ahadic and \Pythia settings were not the same: only $69$ observations were shared between the complete set of outputs considered for their HM waves.

The set of basis functions $h_i(x)$ for each output was comprised of polynomial basis functions up to quadratic order (including cross-terms between different input parameters), similar to the individual-bin emulators in \textsc{Professor}, and for any given output this fairly large candidate set was reduced with the aim of balancing explanatory power and model complexity. The prior variance $\sigma^2$ for each output was taken to be the residual standard error of the fitted polynomial; in effect, we consider the weakly stationary process $u(x)$ to represent the residual surface.
The outputs were seen to vary smoothly with respect to varying input parameter values, motivating the use of a commensurately smooth correlation function 
\begin{equation*}
    c(x,x^\prime)=\exp^{-(x-x^\prime)^2/\theta^2}   
\end{equation*}
with $\theta$ the correlation length, its value chosen in line with the so-called `Durham heuristic' \cite{bower2010galaxy}. This states that, for an emulator whose basis functions are no higher than order $k$, the correlation length should be bounded by the theoretical maximum and minimum distances between the roots of an order $k+1$ polynomial. Given this constraint, we obtain a bounded maximum \emph{a posteriori} (MAP) estimate of $\theta$ for each emulator.

Finally, any parameters which did not singly contribute to the basis functions for an output (that is, a parameter $x^{(i)}$ for which no basis function $h_k(x)$ has the form $x^{(i)}$ or $x^{(i)2}$; combinations $h_k(x) = x^{(i)}x^{(j)}$ for $i\neq j$ may still be present) were considered inactive, and we specified the nugget term by demanding $\Var[w(x)]=\delta\sigma^2$, with $\delta\in[0, 0.05]$ determined during the fitting process in the same MAP estimation used to determine $\theta$.
Correspondingly, we set the variance $\Var[u(x)]=(1-\delta)\sigma^2$, to ensure that the overall residual variation agrees with our prior specification~\footnote{We note here that, while less onerous than those required for a full Gaussian Process emulator, the prior specifications required here still demand a degree of expert elicitation coupled with statistical knowledge. For the former, there is no substitute for domain experts; for the latter, we may (and do) leverage the tools available in the \texttt{hmer} package \cite{iskauskas2021hmer}.}.

\begin{figure}[!h]
    \centering
    \includegraphics[width=0.9\textwidth]{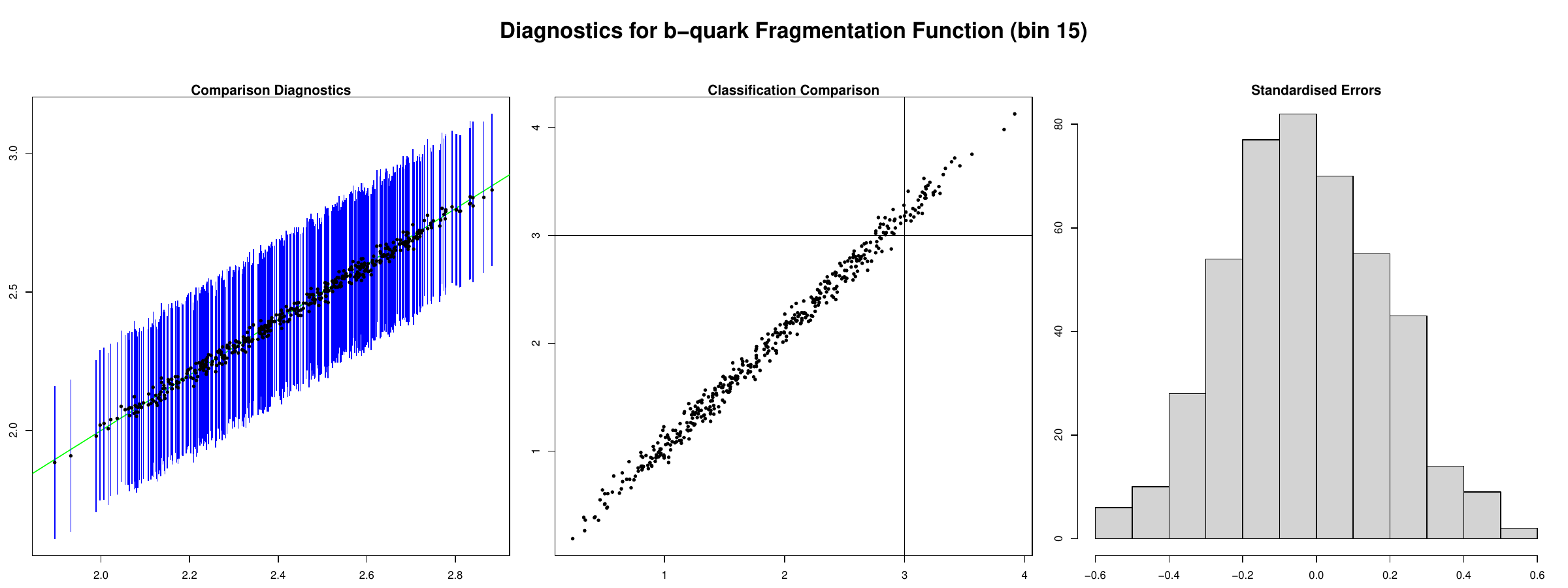}
    \caption{\small Plots of the diagnostics performed on each emulator at each HM wave, for a late-wave \Pythia output. From left to right, the diagnostics check predictive agreement between simulator ($x$-axis) and emulator ($y$-axis) output; implausibility classifications for simulator and emulator; and standardised prediction errors for the emulator. The top right quadrant of the central plot also indicates the proportion of space we expect this emulator to cut out, based on the proportion of validation points lying in this quadrant. Any diagnostic failures would be highlighted in red, rather than blue/black.}
    \label{fig:diageg}
\end{figure}

For each HM wave for a given physics model we considered $800$ parameter sets drawn from the non-implausible parameter-space volume for which we produced simulator outputs. Emulators were trained on the \Ahadic output using $300$ parameter sets and on the \Pythia output using $350$ parameter sets.
The increase in size of the training data for \Pythia stemmed from the higher dimensionality of the input space, as mentioned previously. 
In each case, any remaining simulator runs at that wave were held out for validation diagnostics. 
Diagnostics performed tested for compatibility between emulator and simulator results, checked for systematic bias in emulator predictions, and ensured that no parameter combination that would not be ruled out by the simulator alone was ruled out by the corresponding emulator. 
A full description of the diagnostic measures can be found in~\cite{iskauskas2022emulation}, and an example of the output for each emulator is shown in Figure~\ref{fig:diageg}. 
The vast majority of the emulators demonstrated diagnostics similar to that presented in Figure~\ref{fig:diageg}, with good agreement between emulator predictions and simulator output, a consistent categorisation of implausibility, and appropriately distributed standardised errors. 
At each wave, however, a small handful of emulators required minor modifications to their prior structure: most often by slightly inflating the prior uncertainty $\sigma^2$, or by the introduction of one or two additional training points.

\subsection{Investigating Emulated Model Behaviour}
Alongside contributing to the HM process via the implausibility measure given by Eq.~\eqref{eq:impproper}, one may also use the emulators as surrogates for the simulator output across the parameter space. 
We may investigate which parameters are most influential for particular model outputs (for instance, determining the size of the peak in the histogram of a given observable) and what the nature of the dependencies between inputs and outputs are; due to the emulators' computational efficiency, we can also plot the expected output behaviour across the space via the predictive expectation $\Exp_D[g(x)]$ and identify regions of parameter space for which more data would be useful, with the emulator variance $\Var_D[g(x)]$ acting as an indicator.

\begin{figure}[!h]
    \centering
    \includegraphics[width=0.8\textwidth]{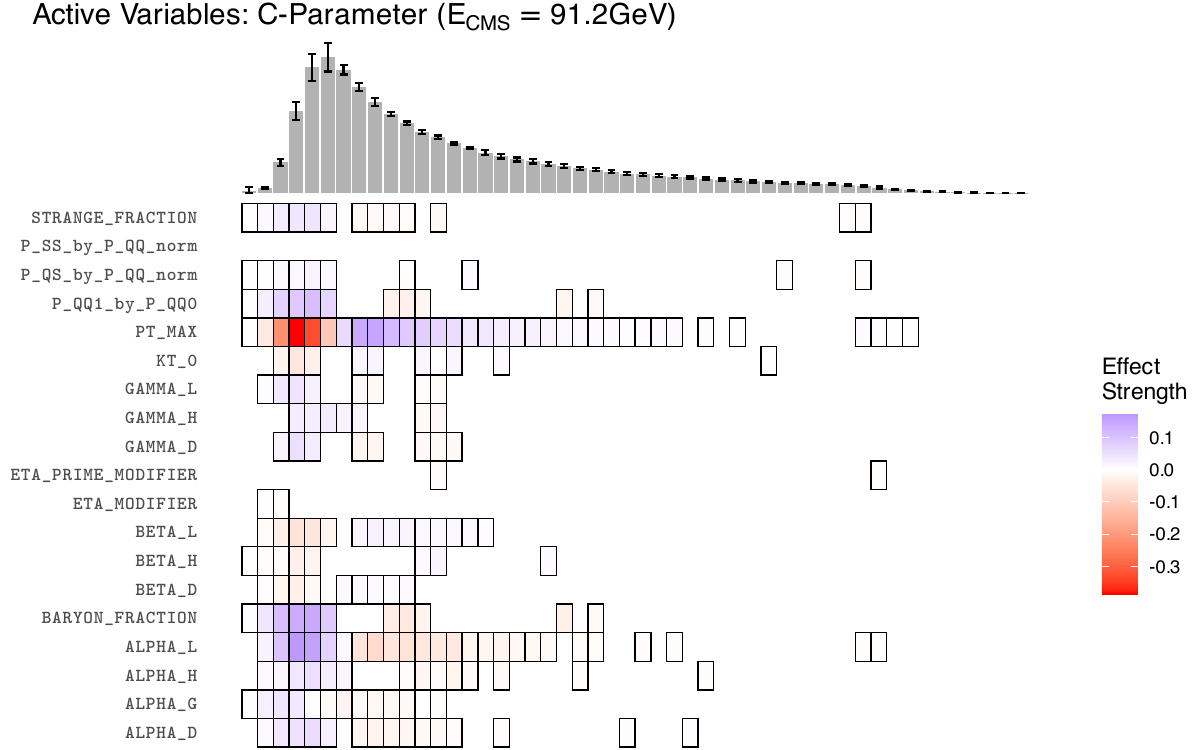}
    \caption{\small The active variables, and strength of effect, for all observable outputs from an ALEPH measurement of the $C$-parameter event shape~\cite{ALEPH:2003obs} at the final HM wave. 
    Any tile with a border indicates that the parameter is active for that output; the colour of the tile determines the strength of the linear effect with deeper blue (red) shades indicating a stronger positive (negative) output response to the parameter. The observed data is overlaid above as a histogram.}
    \label{fig:alephactive}
\end{figure}

Figure~\ref{fig:alephactive} gives a broad overview, and acts as a sense-check, of the output dependence on parameters. 
For this illustration, we consider the final HM wave of the \Ahadic model with data from the ALEPH experiment for the $C$-parameter event-shape distribution~\cite{ALEPH:2003obs}. 
An outlined box in the $(i,j)^{th}$ position in the plot indicates that parameter $x^{(j)}$ is active for output $f_i(x)$, offering an at-a-glance indication of active variable clustering across outputs. 
Beyond the binary classification of active/inactive for each variable and each output, we might also consider the scale of the effect that each active variable has -- this is provided by the colouration of each cell. 
We see that at this late stage of HM, many parameters contribute to the outputs, as one would expect: having reduced the parameter space by a handful of HM waves, the change in the output response is more sensitive to varying parameters than in the original, unrestricted, space where large-scale dependence dominates. 
However, even at this point in the HM process, we see that some parameters are simply not informative for this particular observable. 
As expected, the \texttt{ETA\_xx\_MODIFIER} parameters have little to no impact on the model outputs (since they only alter the probability for the production of $\eta$ mesons). 
A similar consideration applies to the outputs, too.
We may note that the final handful of outputs are barely affected by changing input parameters. 
This is, in some sense, unsurprising. 
We expected a high correlation between outputs, and since these active variable determinations arise from the final wave of simulator runs, restricting the non-implausible space by using any of the outputs from this analysis has all but completely restricted the possible output values of the tail of this histogram. 

We detect similar structures in the colouring of the grid elements, and we realise that varying particular parameters dominates the model response: 
for example, the \texttt{PT\_MAX} and \texttt{ALPHA\_x} parameters seem to drive much of the output changes. 
We also notice an interesting, but common, dependence: increasing the \texttt{ALPHA\_x} parameters has a positive effect on the earlier bins of the histogram and the reverse effect on those later bins. 
This agrees with physical intuition, as the change in behaviour corresponds to the areas before and behind the peak of the histogram, and an increase in these parameters forces the peak of the histogram both to increase in scale but also to move it further to the left, depressing any bins later in the histogram.

The information exemplified in Figure~\ref{fig:alephactive} provides first insights into how the model reacts to different parameter combinations. 
Such considerations could feed into justifications for dimensional reduction of the input space. 
For instance, if our only interest was in the observations of this particular observable we might argue that much of the model response could be captured by half of the parameters, and modify our model accordingly~\footnote{
    We would, of course, include additional model discrepancy to account for the omission of these parameters. 
    Happily, we could estimate this extra output variability due to varying the omitted parameters with recourse to the emulators, too.}. 
For models with many parameters and an extremely high computational load, we might also use this information to partition our parameter space into a `border-block' structure, clustering outputs into collections determined by their active variables and proposing new points based on a Cartesian product of proposals from these active variable subsets~\cite{cumming2009small}. 
This is similar in spirit to other methods that have been used to cluster parameters (see, for example, \cite{Bellm:2019owc}). 
The method proposed here, however, does not require us to specify some goodness-of-fit to test parametric clusters. 
Since the determinations of active variables are made by the emulators, it is independent of the particular simulator runs we have available to us and hence of any sampling of a statistical measure of fit. 
Furthermore, it allows us to partition both in the parameter and output spaces simultaneously, since we obtain general active variable subsets for each output.

\begin{figure}[!h]
    \centering
    \begin{subfigure}{0.49\textwidth}
        \includegraphics[width=\textwidth]{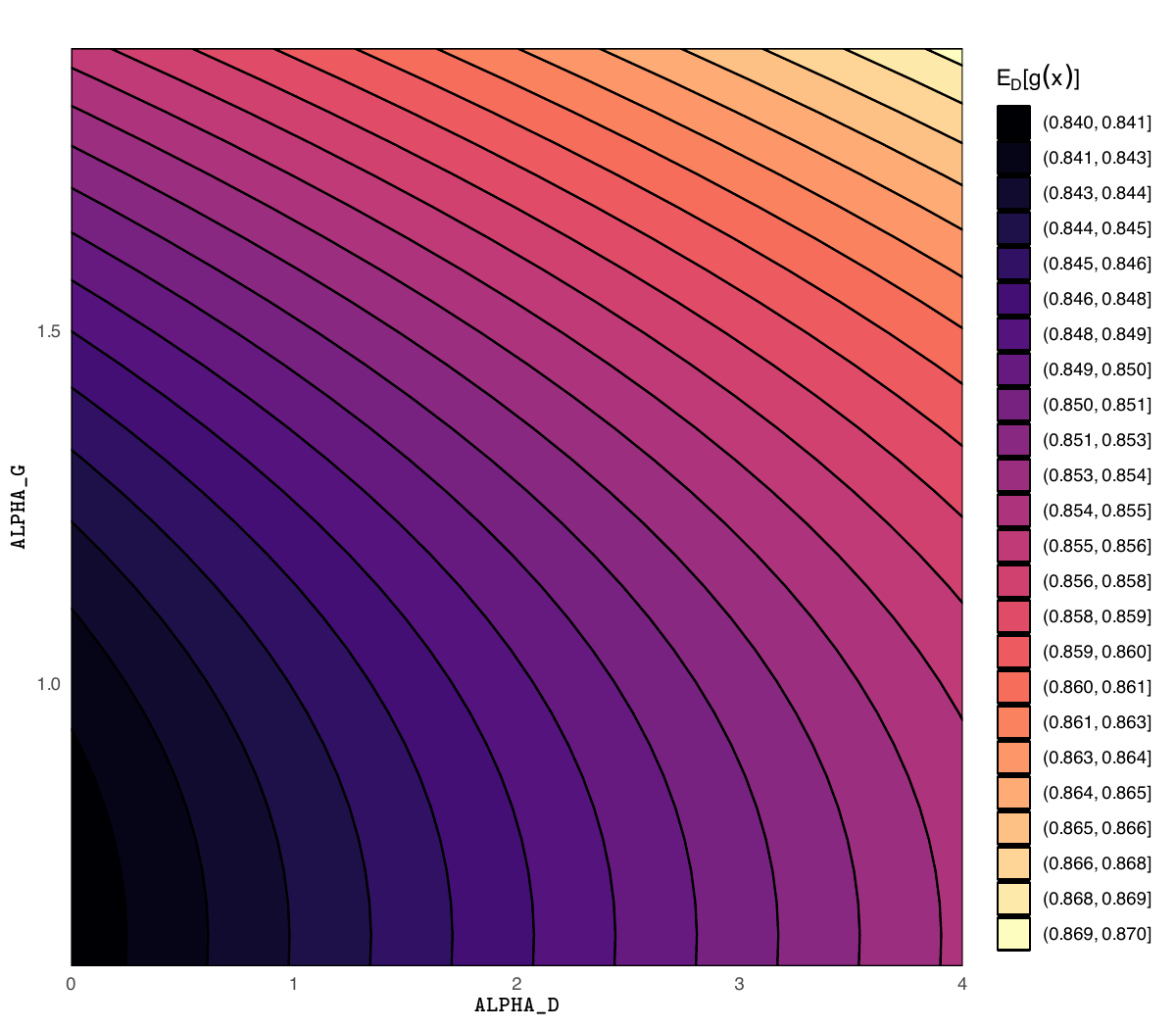}
    \end{subfigure}
    \begin{subfigure}{0.49\textwidth}
        \includegraphics[width=\textwidth]{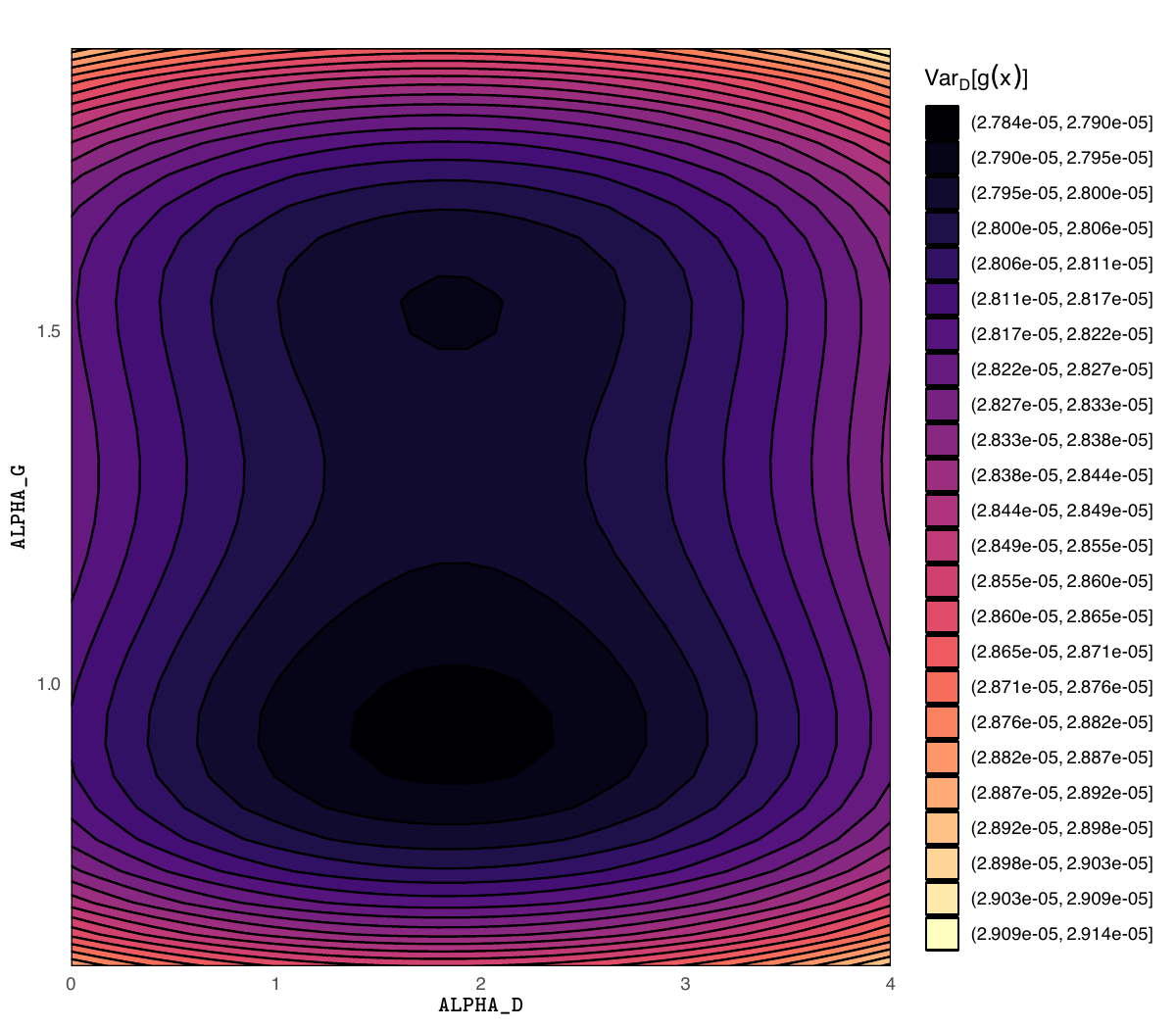}
    \end{subfigure}
    \caption{\small The emulator predictions (left) and corresponding uncertainty (right) for an output from DELPHI \cite{DELPHI:1996sen} at wave $1$, plotted with respect to two of the \Ahadic parameters. All other parameters are held at their mid-range values.}
    \label{fig:emexpvar}
\end{figure}

For any given combination of parameters, we may also investigate the model behaviour across the space through the trained emulators and without direct recourse to the computationally expensive simulator. 
Figure~\ref{fig:emexpvar} shows the predicted simulator output $\Exp_D[g(x)]$ (left) and corresponding emulator uncertainty $\Var_D[g(x)]$ (right) for one of the model outputs, whose form was given in Eqs.~\eqref{eq:ED} and \eqref{eq:VarD}. 
We can see that the broad effect of increasing either of \texttt{ALPHA\_G} or \texttt{ALPHA\_D} results in a general increase in the model response, though the relationship is far from linear. 
The emulator uncertainty also displays interesting characteristics. 
In particular, the general increase in uncertainty at the edges of the space is unsurprising as we are unlikely to have placed any training points on the boundary of the input space.
The emulator's larger uncertainty here will motivate some exploration of the edges of the space via the implausibility measure, as we would expect. 
Evaluating the emulator even at  very fine-grained grids of points is fast, which allows us to produce comprehensive sets of plots showing predictions with respect to any pair of input parameters at various slices through the multi-dimensional input space.
Such predictions offer greater insight into the model behaviour, or serve as a sense-check for our physical intuition. 
They also help to identify any potential conflicts between our model and observational data, without requiring a large amount of computational time.

Before turning to the results, we briefly comment on one final consideration in the HM process. 
As mentioned in  Section~\ref{sec:hme}, we proceed through HM waves until a predetermined stopping criterion has been met. 
In this setting, we deemed the process to have terminated if any of the following conditions were met:
\begin{itemize}
    \item A high proportion ($>90\%$) of the outputs emulated at the current wave had low emulator uncertainty, or had been previously emulated with low uncertainty: training further emulators would only reduce this emulator uncertainty, already subdominant to the other terms in the implausibility, and therefore provide little inferential or predictive knowledge;
    \item The non-implausible input parameter space and output space is stable: if the current collection of emulators removes less than $5\%$ of the previous volume, and the output ranges similarly showed no substantial difference between this and the previous non-implausible region, then further HM would have little benefit for the added computational cost~\footnote{One might view this as searching for an `elbow' in the volume of non-implausible parameter space over the HM waves.};
    \item The non-implausible space is empty;
    \item A suitably large proportion of the proposed points give rise to model runs with high agreement to the observed data. While we did not expect to match to every target~\footnote{This naturally follows from consideration of the observations themselves. 
    If we assume that each is uni-modally distributed around the observational value with standard deviation corresponding to the observational error, and we assume that \emph{all} bins of an analysis's histogram are mutually correlated, then the probability of the truth lying within the observational windows of all $29$ observables is less than $25\%$.}, we anticipated matching to a high number of them: around $400$ of the $432$ observations would be considered a good fit.
\end{itemize}
We note that the choice of stopping conditions and the specific numerical values used here are application-specific and (to an extent) pragmatic, but we argue that these conditions are appropriate to other applications with minor modifications.
For instance, if one's ultimate aim is to use the final-wave emulators as a true surrogate for the model, then we would demand a higher proportion (possibly all) of the emulators to have low emulator uncertainty; in circumstances where we have fewer or less highly-correlated outputs, it would be reasonable to demand that all outputs are matched by simulator runs.

Applying these stopping conditions resulted in the termination of the \Ahadic (\Pythia) history match after $3$ ($5$) full waves.
At no stage did the emulators, or the non-implausible space generated, belie the appropriateness of our prior specifications: happily, at no stage was the non-implausible space deemed empty~\footnote{Had this been the case, one might have considered whether we had severely underestimated the model discrepancy, necessitating a restart of the process.}.
The requirement of additional HM waves for \Pythia is unsurprising, considering the higher dimensionality of the input parameter space and the wider initial ranges for those parameters, as shown in Table~\ref{tab:params_pythia}.

\section{Calibration Results for \Ahadic and \Pythia}\label{sec:results}
The final outcome of the HM calibration is given by a set of non-implausible parameter points that can be used to estimate the hadronisation models' internal uncertainties. 
Doing this consistently for the \Ahadic and \Pythia fragmentation models, we can furthermore derive a combined non-perturbative uncertainty that also reflects aspects of physics-model dependence.

In the following, for both \Ahadic and \Pythia, we will examine two complementary aspects: first, the structure of the surviving parameter space, its volume and correlations, and second, the spread in predictions these parameter sets induce to physical observables.

\subsection{Non-Implausible Parameter Space}
Understanding the structure of the final, non-implausible parameter space is obviously of great importance. 
The volume of the surviving space provides a measure of how well the data constrain the model, while its topology and shape allow us to draw conclusions about parameter correlations that are impossible to analyse based on physical observables alone. 
In particular, large connected regions indicate directions in parameter space along which variations lead to small changes in the predicted observables. 
In contrast, disjoint areas indicate alternative solutions which are notoriously difficult to explore with MCMC- or ABC-based algorithms \cite{mckinley2009inference, geyer2011introduction}.

We begin by summarising how the non-implausible volume evolves across the successive waves in our calibration of the two models. 
The parameter ranges initially considered, defining the initial hyper-rectangular parameter-space volume, are listed in Tables~\ref{tab:params_ahadic} and \ref{tab:params_pythia} in Appendix~\ref{app:parameter_selection} for \Ahadic and \Pythia, respectively. 
To estimate the non-implausible parameter-space volume after a HM wave, we compute the bounding box of the $800$ wave members.
However, this should be interpreted as indicative rather than precise measures of the non-implausible region. 
For \Ahadic, the first wave reduces the estimated parameter-space volume by roughly three orders of magnitude (from $1.0$ to $3.0\times10^{-3}$); the volume continues to decrease over the subsequent two waves to $7.7\times10^{-4}$ and $5.2\times10^{-4}$.
For \Pythia, the first wave removes nearly two orders of magnitude of the initial volume (from $1.0$ to $2.2\times10^{-2}$), and the estimated volume continues to shrink over the remaining five waves, reaching $3.4\times10^{-3}$, $1.1\times10^{-5}$, $2.2\times10^{-6}$, $8.0\times10^{-9}$, and finally $1.5\times10^{-13}$, amounting to a total reduction of over thirteen orders of magnitude.
At their respective late HM stages, the majority ($>80\%$) of emulators of both models had low uncertainty, and the corresponding simulator runs were matching to at least $390$ of the observations. 
While each of these results, taken in isolation, do not satisfy any of our stated stopping conditions, their combination is highly suggestive that no value would be gained from performing additional simulator runs and training new emulators to the results.
As a final check, we trained an additional wave of emulators to each of these `final' simulator runs to propose an extra wave of non-implausible points -- the volume of the non-implausible regions thus obtained showed little reduction.
This additional wave for \Ahadic gave a non-implausible volume of $4.0\times 10^{-4}$, and that for \Pythia a volume of $1.3\times 10^{-13}$, representing a space reduction of around $23\%$ and $13\%$, respectively.
These results, taken together, were sufficient to judge that in each case the continuation of the HM process would provide no material improvement to justify the additional computational expense. It is worth noting that we have not, strictly speaking, satisfied any individual one of our pre-determined stopping conditions; rather, our closeness to the satisfaction of multiple of them is considered here to be sufficient evidence that further waves would be unnecessary.

\begin{figure}[h!]
    \centering
    \includegraphics[width=0.49\linewidth]{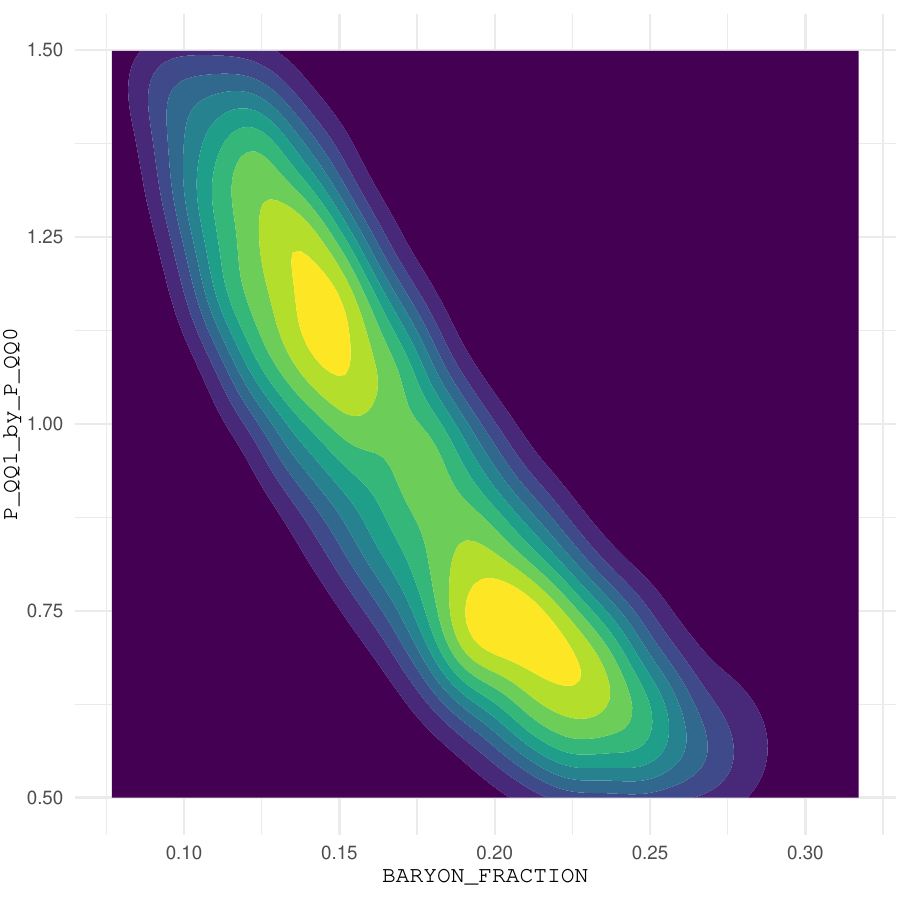}
    \includegraphics[width=0.49\linewidth]{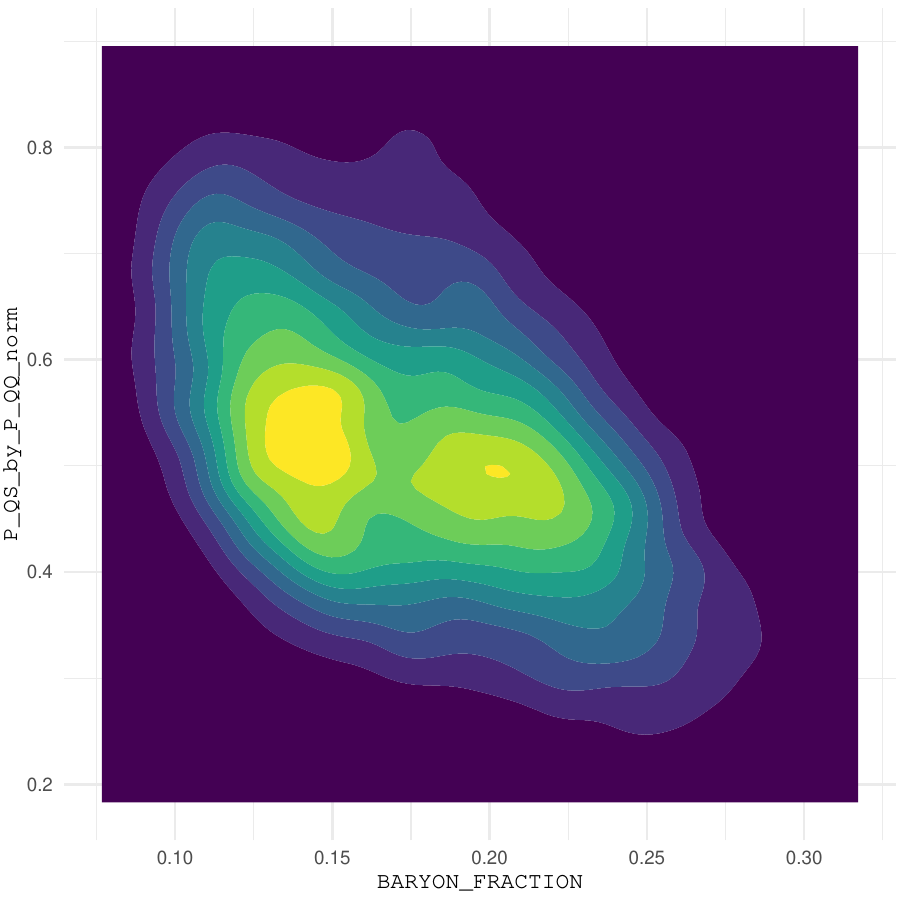}
    \caption{
    Selected two-dimensional projections of the non-implausible \Ahadic parameter space for the final HM wave.
    The densities are estimated from $10^4$ sampled points and normalised such that each contour region represents an increment of $10\%$ probability. The full collection of two-dimensional projections can be found in Appendix~\ref{app:additional_figures}.
    }
    \label{fig:parameter_space}
\end{figure}

The residual non-implausible space exhibits a quite complex structure for \Ahadic and \Pythia, illustrated by exemplary two-dimensional projections of the parameter space. 
Figure~\ref{fig:parameter_space} presents two such projections across the final wave of HM for \Ahadic, focusing on correlations between parameters that govern the non-perturbative flavour production in gluon splittings and cluster decays. 
The \texttt{BARYON\_FRACTION} parameter (initial range $[0.0,0.7]$) steers the overall production rate of di-quark pairs, while \texttt{P\_QQ1\_by\_P\_QQ0} (initial range $[0.5,1.5]$) affects the ratio of di-quarks in the spin-$1$ and spin-$0$ states, and \texttt{P\_QS\_by\_P\_QQ\_norm} (initial range $[0.1,0.9]$) selects di-quarks with or without strange-quark content~\cite{Chahal:2022rid}. 

After the final HM wave the non-implausible space for these parameters is indeed significantly reduced, compared to the initial parameter ranges. 
Unsurprisingly, the values of the \texttt{BARYON\_FRACTION} and the \texttt{P\_QQ1\_by\_P\_QQ0} parameter are clearly correlated. 
In the final wave, the sampled points approximately form a `banana-shaped' distribution with two distinct maxima at opposite ends. 
A similar bimodal structure is observed for the two-dimensional projection on \texttt{BARYON\_FRACTION} versus \texttt{P\_QS\_by\_P\_QQ\_norm}, although in this case the two peaks are aligned at similar values  of \texttt{P\_QS\_by\_P\_QQ\_norm}. 
Similar bi-modal and multi-modal structures and non-trivial correlations are indeed observed for a variety of parameter pairs, see Figure~\ref{fig:2d_ahadic_correlations} in Appendix~\ref{app:additional_figures}. 
These findings highlight the strength of the history-matching approach and illustrate the limitations of single-point calibration approaches: a tune that settles in one of the two modes would potentially miss the equally viable region around the other.

\begin{figure}[h!]
    \centering
    \includegraphics[trim={0 0 0 0},clip,width=0.49\linewidth]{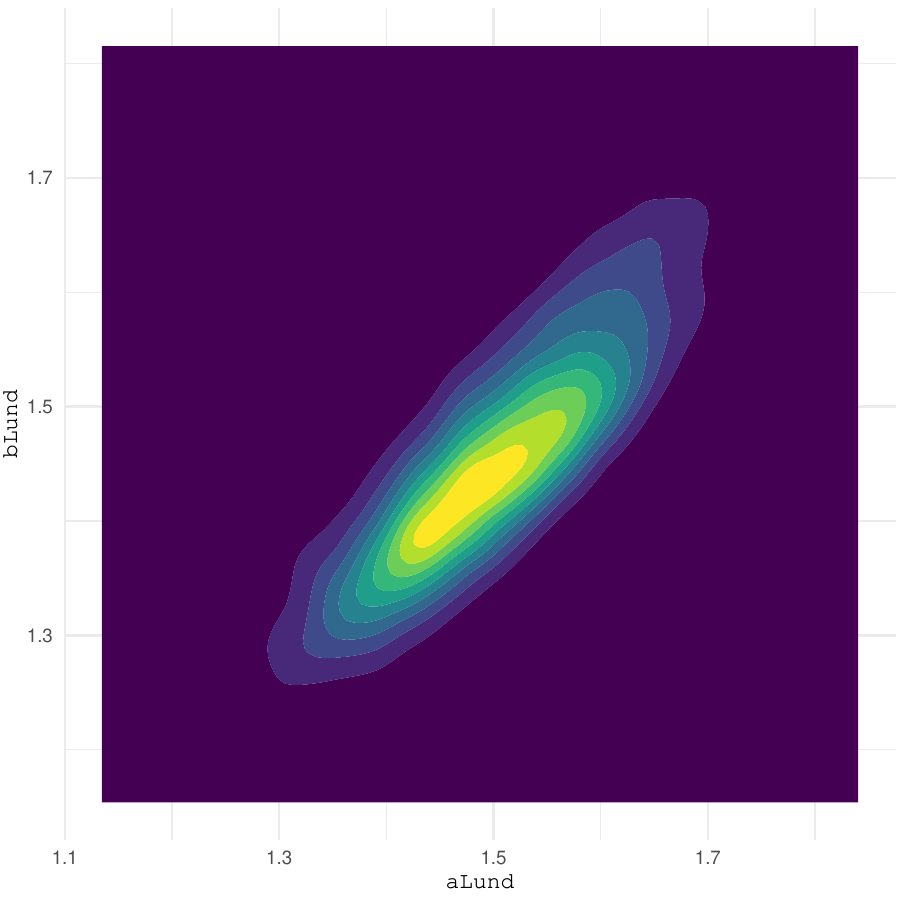}
    \includegraphics[trim={0 0 0 0},clip,width=0.49\linewidth]{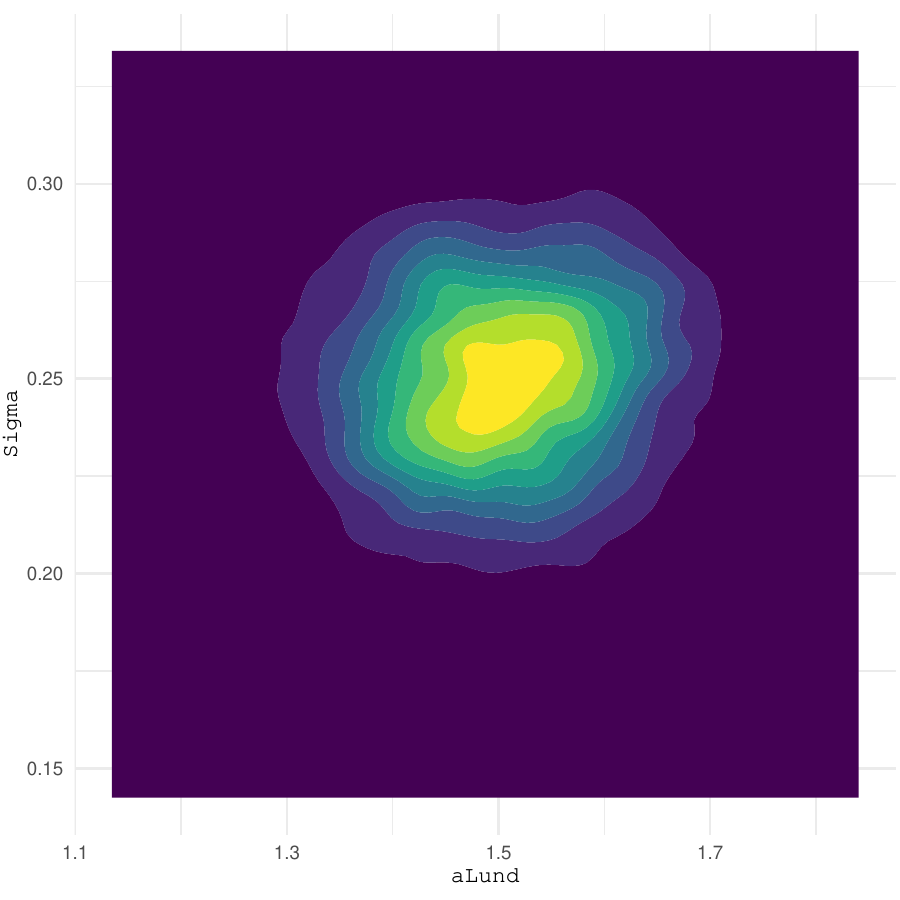}
    \caption{Selected two-dimensional projections of the non-implausible \Pythia parameter space for the final history-matching wave. 
    The densities are estimated from $10^4$ sampled points and normalised such that each contour region represents an increment of $10\%$ probability. The full collection of two-dimensional projections can be found in Appendix~\ref{app:additional_figures}.}
    \label{fig:parameter_space_pythia}
\end{figure}

A qualitatively similar picture emerges for \Pythia, where the correlations among the string-fragmentation parameters are equally rich (see Figure~\ref{fig:2d_pythia_correlations} in Appendix~\ref{app:additional_figures} for a complete overview). 
In Figure~\ref{fig:parameter_space_pythia} we show two examples that involve the parameter $\sigma$ (\texttt{Sigma}, initial range $[0.0,1.0]$), and the $a$ (\texttt{aLund}, initial range $[0.0,2.0]$) and  $b$ (\texttt{bLund}, initial range $[0.0, 2.0]$) parameters of the Lund symmetric fragmentation function~\cite{Andersson:1983ia}. 

The left panel of Figure~\ref{fig:parameter_space_pythia} confirms the well-known feature that parameters \texttt{aLund} and \texttt{bLund} are indeed strongly correlated. 
For the projection on the \texttt{aLund}--\texttt{Sigma} plane, shown in the right panel, we observe a unimodal distribution that, in fact, is quite tightly constrained by the data.  

\subsection{Observable Predictions and Model-Parameter Sensitivities}
While the parameter-space volume and correlations characterise how strongly the model is constrained internally, the impact of these constraints on measurable quantities is our main phenomenological interest.
By propagating the remaining plausible parameter sets through \Sherpa and the corresponding hadronisation models, we can quantify the resulting spread in the predictions of physical observables and interpret it as a non-perturbative uncertainty. 
Furthermore, this procedure allows us to assess which observables are robust against variations in the hadronisation model and which exhibit sensitivity to regions of parameter space that were unconstrained by the data used for calibration. 
Hypothetically, information like this can also be used to devise new observables or measurements that help to pinpoint the parameters of a given model or even uncover modelling deficiencies.  

\begin{figure}[h!]
    \centering
    \includegraphics[width=0.49\linewidth]{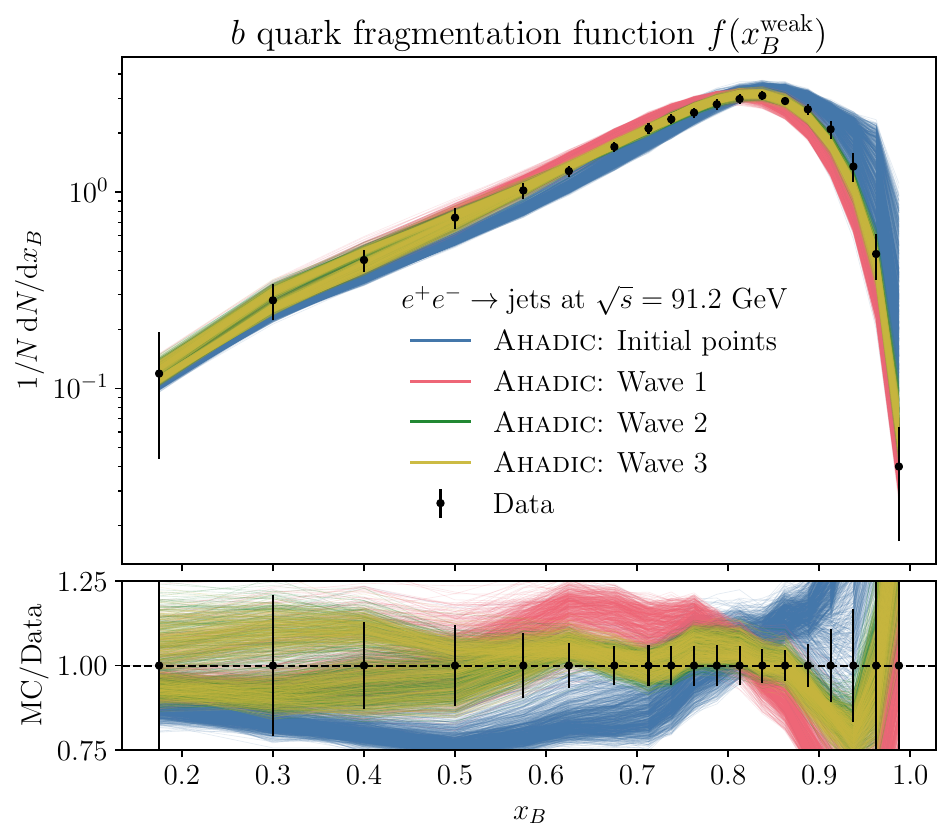}
    \includegraphics[width=0.49\linewidth]{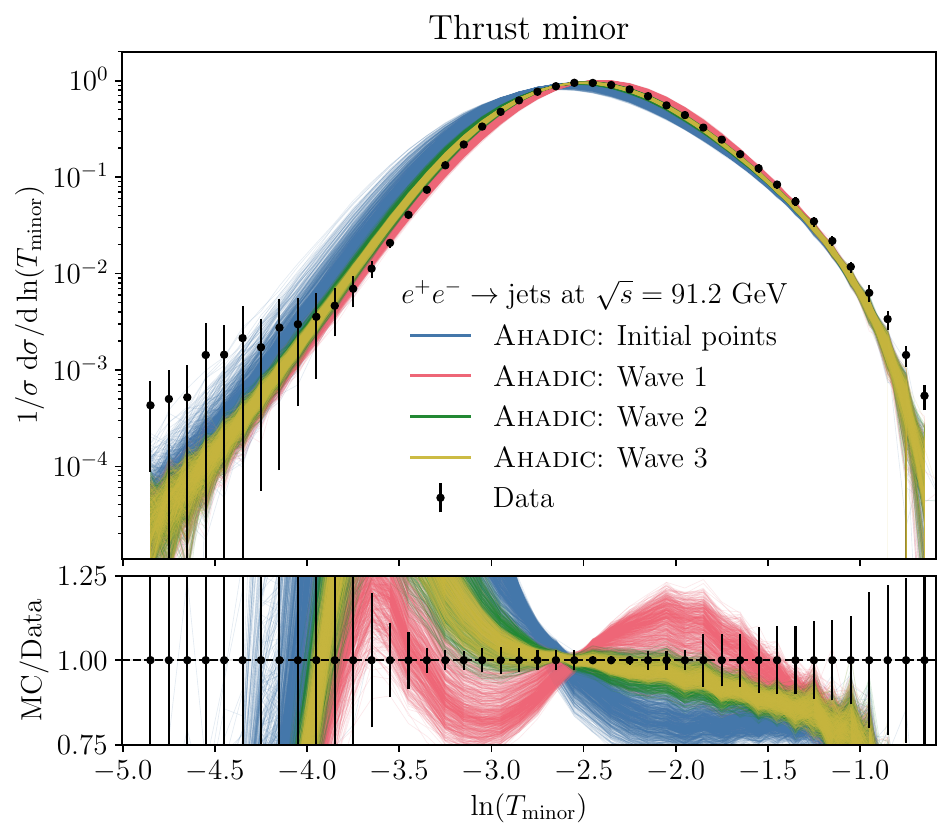}
    \caption{
    Observable envelopes for the sequence of \Ahadic history-matching waves. 
    Left: the $b$-quark fragmentation function with experimental data from~\cite{ALEPH:2001pfo}. Right: the thrust-minor event shape with experimental data from~\cite{ALEPH:2003obs}.
    }
    \label{fig:observables_all_waves}
\end{figure}

Figure~\ref{fig:observables_all_waves} shows the envelope formed by all \Sherpa runs for two selected observables, the $b$-quark fragmentation function (left panel) and the thrust-minor event-shape distribution (right panel), both as measured by ALEPH at LEP~\cite{ALEPH:2001pfo,ALEPH:2003obs}, across the sequence of HM waves for \Ahadic.  
Since the parameter space of each wave is a strict subspace of that of the previous wave, we expect the envelope to shrink for successive waves and to remain contained within the one of the preceding wave.
This is indeed borne out by the simulation data: the second wave already converges much more closely to the experimental measurements, and the final wave's envelope is, for most of the observable range, comparable in size to the experimental uncertainties, representing our estimate of the non-perturbative uncertainty of the \Ahadic hadronisation.

Interestingly, the observable spread can \emph{increase} after the first wave despite the parameter-space volume having already decreased. 
Since our space-filling design is, by necessity, comprised of a small number of parameter combinations, we do not anticipate that the full scope of model output behaviour is demonstrated by the corresponding simulator evaluations, and the first wave of HM can explore previously unsampled regions that turn out to produce new observable behaviour.
Rather than excluding these regions, HM identifies them as particularly informative and prioritises them for subsequent waves.
This behaviour is visible for both observables in Figure~\ref{fig:observables_all_waves}, where the envelope of the initial design is not a strict superset of that of the first wave.

\begin{figure}[h!]
    \centering
    \includegraphics[width=0.32\linewidth]{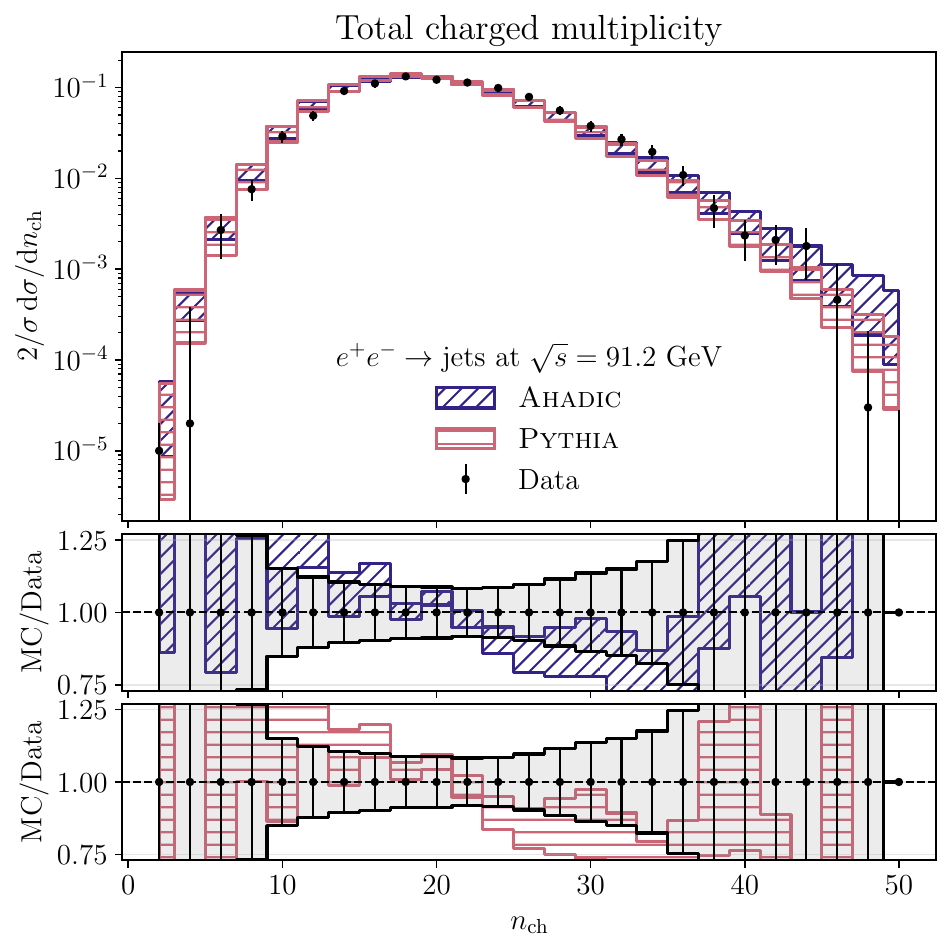}
    \includegraphics[width=0.32\linewidth]{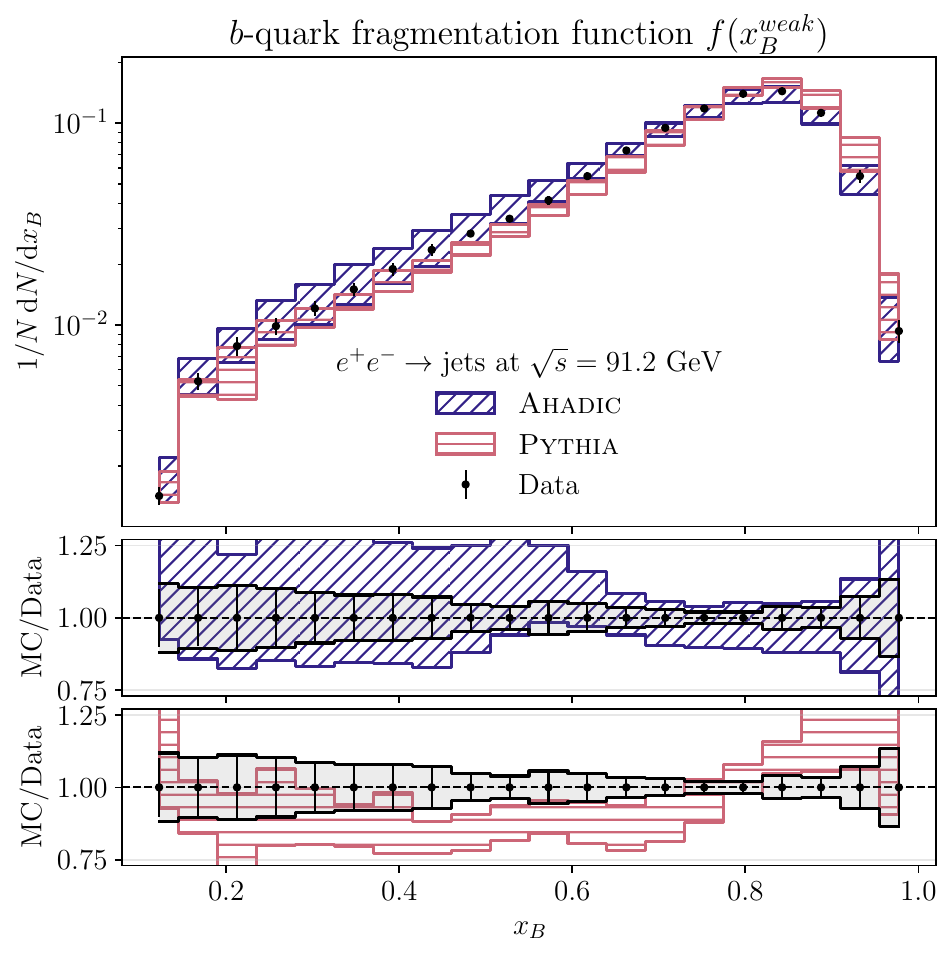}
   \includegraphics[width=0.32\linewidth]{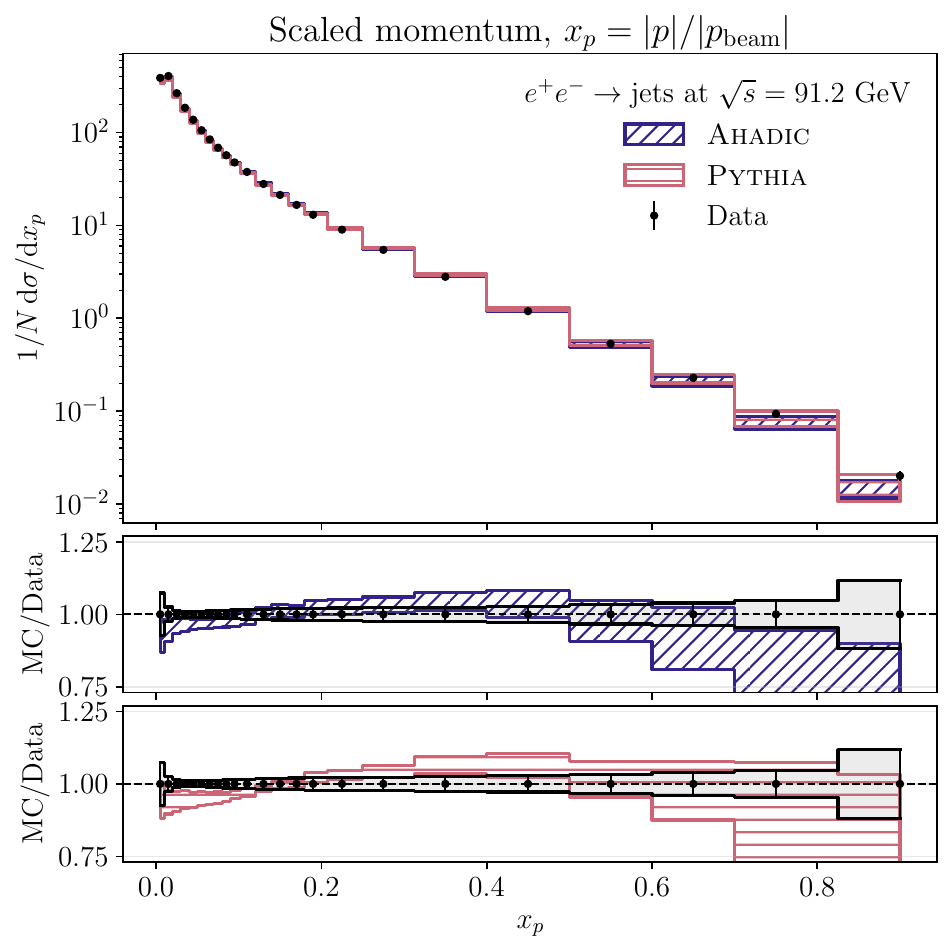}
   \includegraphics[width=0.32\linewidth]{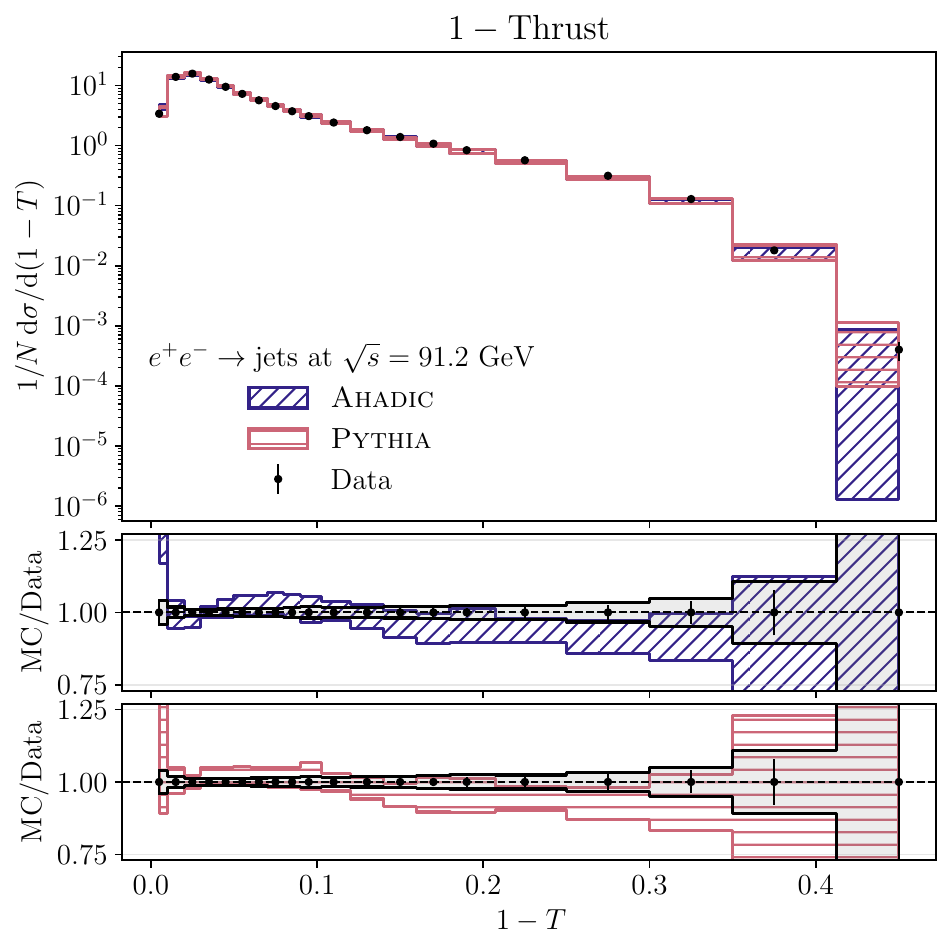}
    \includegraphics[width=0.32\linewidth]{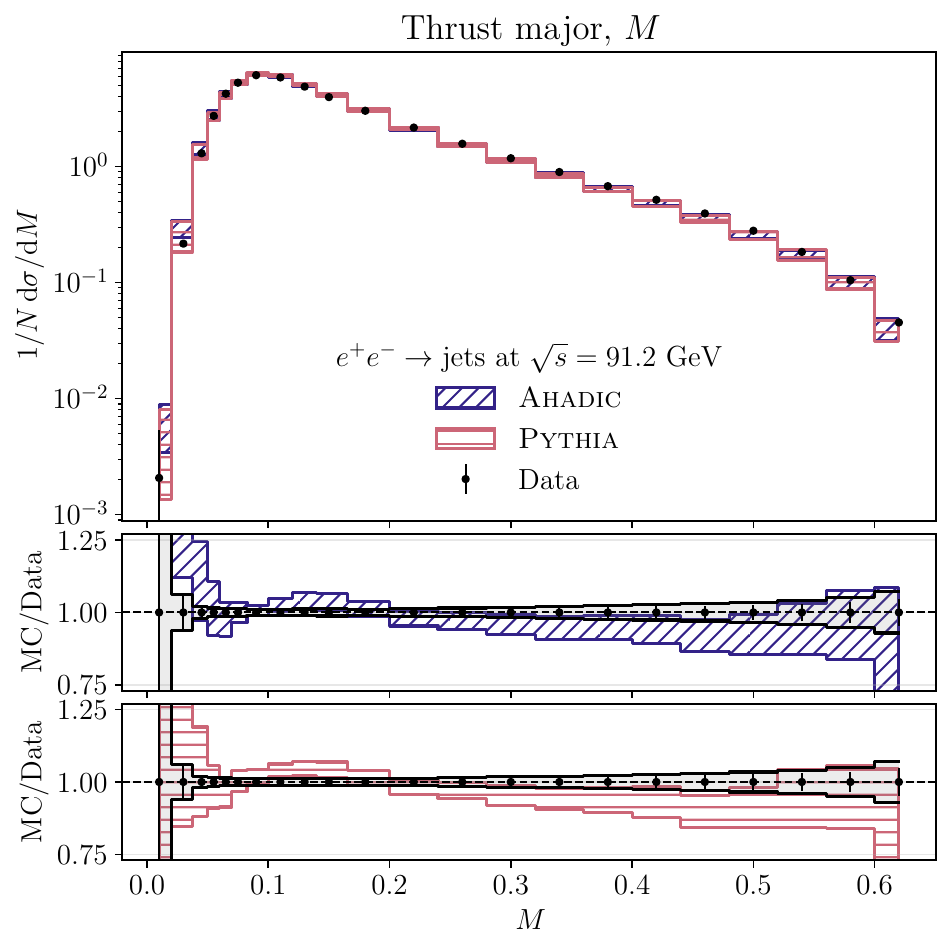}
    \includegraphics[width=0.32\linewidth]{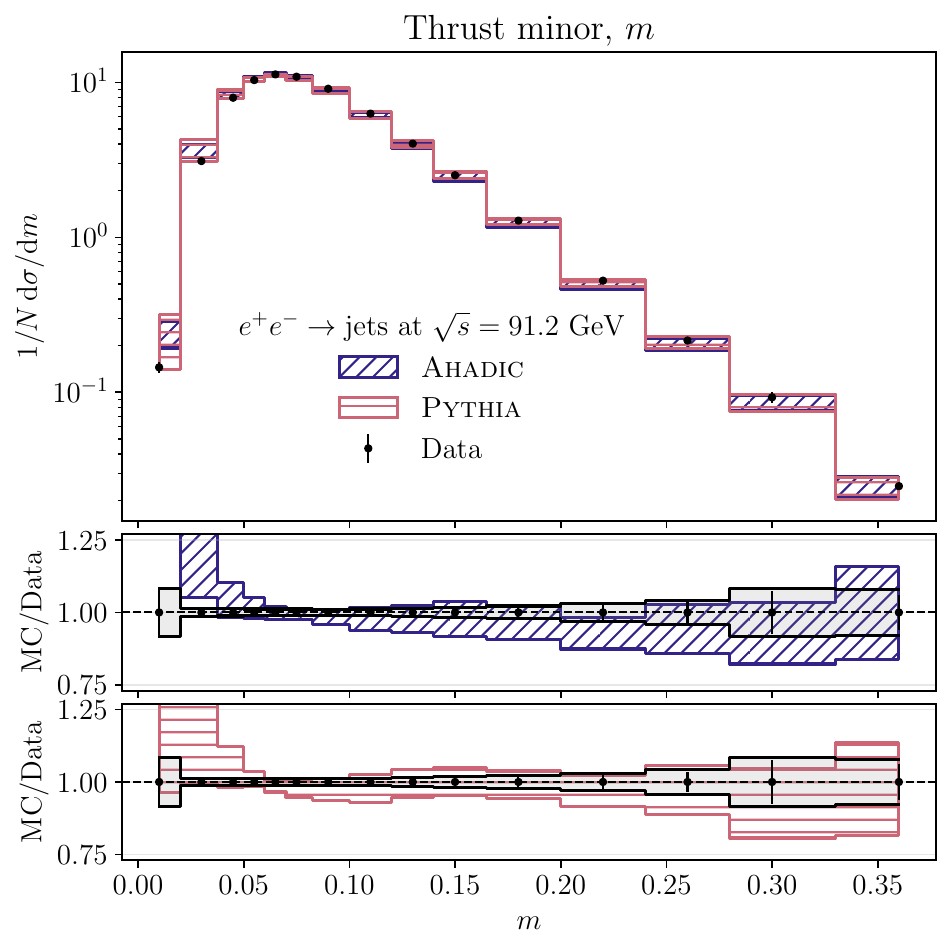}
    \includegraphics[width=0.32\linewidth]{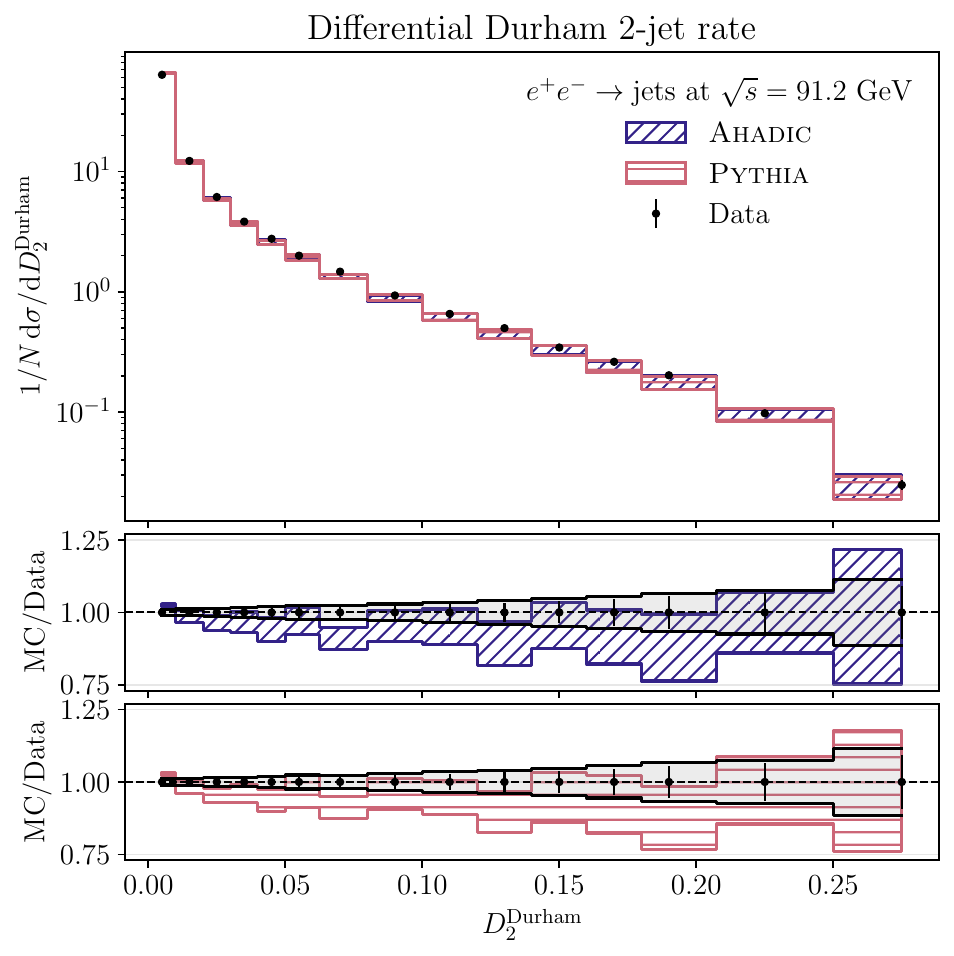}
    \includegraphics[width=0.32\linewidth]{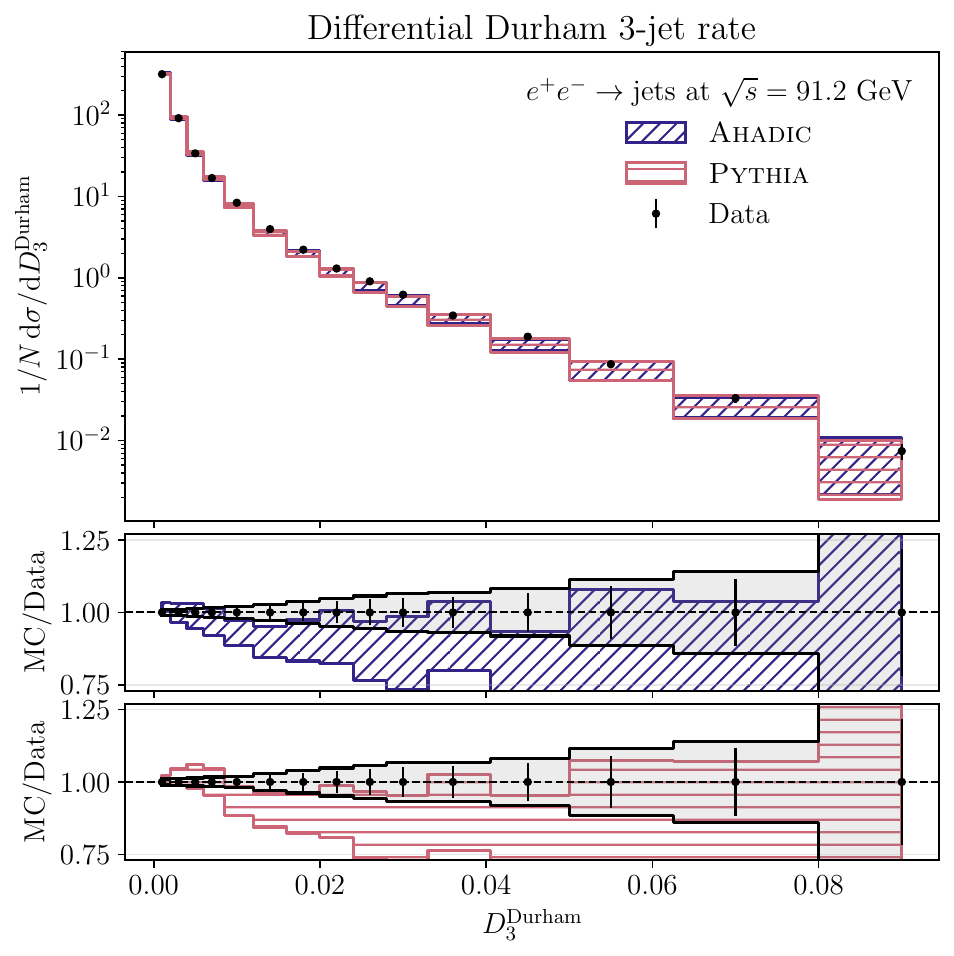}
    \includegraphics[width=0.32\linewidth]{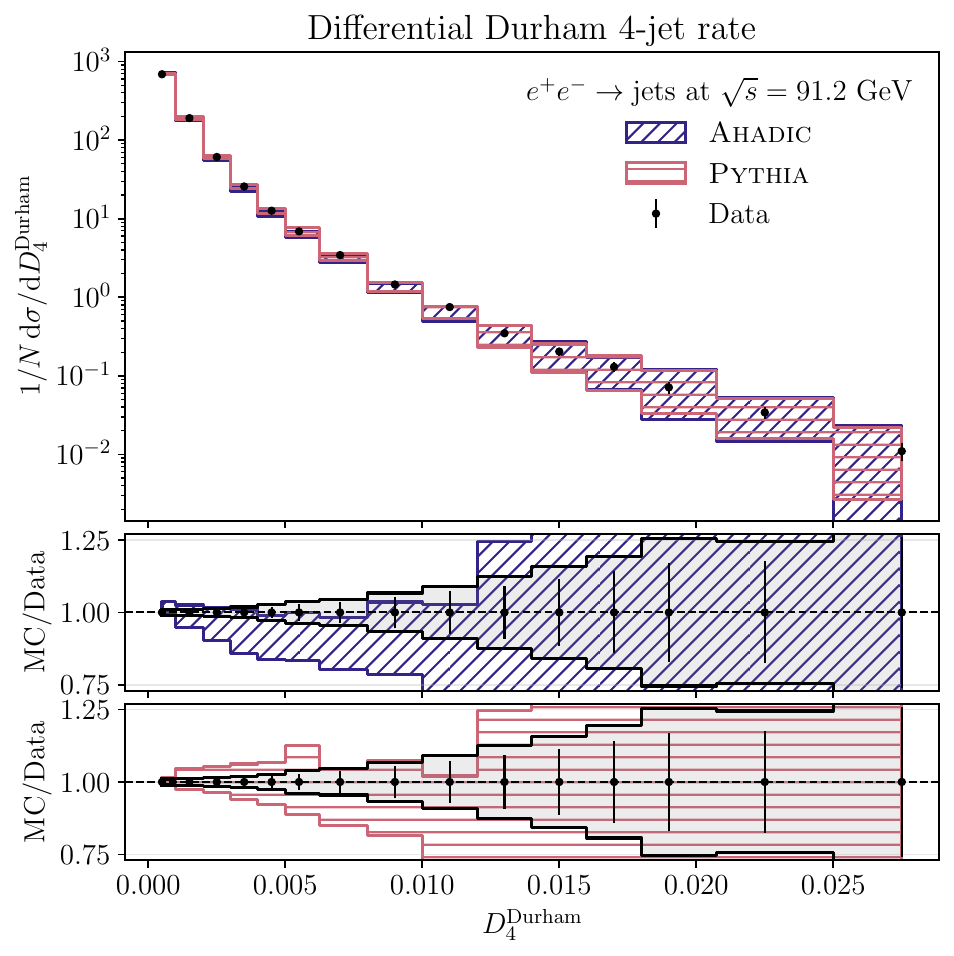}
    \caption{
    Comparison of \Sherpa\!\!+\Ahadic (blue) and \Sherpa\!\!+\Pythia (orange) predictions with experimental data from $e^+e^-$ annihilation at $\sqrt{s}=m_Z$.
    Top row, left to right: charged-particle multiplicity~\cite{ALEPH:1991ldi}, $b$-quark fragmentation function~\cite{OPAL:2002plk}, and charged-particle scaled momentum~\cite{DELPHI:1996sen}.
    Middle row: thrust, thrust major, and thrust minor~\cite{DELPHI:1996sen}.
    Bottom row: differential jet rates $D^\text{Durham}_{2,3,4}$ in the Durham scheme~\cite{DELPHI:1996sen}.
    The coloured shaded bands represent the envelopes spanned by all parameter sets in the final HM wave for each model, while the black band indicates the combined error of the experimental data (black markers) and the average statistical uncertainty of the simulator predictions. 
    }
    \label{fig:physics_observables}
\end{figure}

In the following, we focus on the final remaining non-implausible region and compare the predictions for the cluster hadronisation, after 3 waves of HM, with those for the Lund-string model, after 5 waves. 
To provide a complete picture, we discuss single-hadron observables, event-shape variables, and differential jet rates in Figure~\ref{fig:physics_observables}, as well as identified hadron yields in Figure~\ref{fig:pdg-hadron-multiplicities}. 
More results can be found in Appendix~\ref{app:additional_figures}.

Overall, we find good agreement between both models and the experimental data. 
The top three figures show the total charged-particle multiplicity, the $b$-quark fragmentation function, i.e.\ the scaled energy distribution of weakly decaying $B$-mesons, and the absolute momentum of charged particles scaled by the beam momentum. 
For the charged-particle multiplicity, both hadronisation models yield consistent results, with 
\Ahadic and \Pythia spanning envelopes of similar size in our final HM wave. 
The difference between the two hadronisation models is more striking for the $b$-quark fragmentation function, where \Ahadic generally shows larger variations that contain the data points. 
In comparison, \Pythia is more constrained, with the final set of points slightly undershooting the data in the central region. 
This trend is only partially visible in the scaled momentum, where the two hadronisation models agree except for the region of large $x_p$, where the \Pythia predictions better cover the data within uncertainties.

Eqs.~\eqref{eq:nativeIMP} and \eqref{eq:impproper} show how our implausibility measure, used to exclude parameter points, crucially depends both on the observational error $e$ and the model discrepancy $\epsilon(x)$. 
To indicate these components, we include in the lower panels a black shaded band, corresponding to the square-root of the quadratic sum of the measurement error (given by the markers) and the averaged statistical error of the generator predictions. 
For both \Ahadic and \Pythia, comparing the extent of this band with the envelope of the parameter variations signals the effect that a given observable bin has in constraining the model parameters. 
Accordingly, the comparably small parameter uncertainties for the peak region of the charged-particle multiplicity distribution largely originate from constraints of other observables. 
In contrast, for the $x_p$ distribution, the measurement uncertainty is of the order or smaller than the parameter uncertainties. 

The central three plots focus on global event-shape variables, i.e., thrust, thrust major, and thrust minor, measured on all final-state particles. 
For all of them, the predictions accurately reproduce the experimental data, and the two hadronisation models yield nearly indistinguishable results across most of the observable range. 
Similarly to the $x_p$ distribution, the combined error is typically smaller than the parameter uncertainties. 

Finally, the three plots at the bottom show the differential $2$-, $3$-, and $4$-jet rates, $D^\text{Durham}_{2,3,4}$, measured on all final-state particles, for the Durham jet-clustering algorithm~\cite{Catani:1991hj}. 
Here, both models again describe the data reasonably well and produce consistent predictions. 
For the 2- and 3-jet rates, \Ahadic is on average 10\% below the experimental data, with the upper part of the envelope mostly coinciding with the data points. 
A similar but not quite as pronounced pattern is also visible for the \Sherpa\!\!+\Pythia predictions.
In the tails of the jet-rate distributions, the envelopes spanned by the combined measurement and statistical error are of the size of the parameter variations for the final HM waves for both \Ahadic and \Pythia. 
In fact, in particular for the $3$- and $4$-jet rate the statistical error on the MC runs is non-negligible, limiting their constraining power.

Figure~\ref{fig:pdg-hadron-multiplicities} shows the average per event identified-particle yield for a plethora of mesons and baryons. 
The predictions from both hadronisation models generally reproduce the experimental data from Ref.~\cite{ParticleDataGroup:2008zun} well, with some notable exceptions. 
We find that \Ahadic produces on average roughly 25\% too many $\omega(782)$ and $K^{*0/+}(892)$, while underestimating the $D^+_{s1}$, $J/\psi(1s)$, and $\psi(2s)$ yields by approximately 30\%. 
For all other hadron multiplicities, the envelope either overlaps with the experimental 
data or lies within 10\% of the measured values.
\Pythia exhibits comparable performance, although it overshoots the $D^+$ rate by 30\% and undershoots the $J/\psi(1s)$ yield by a similar amount 
as \Ahadic. 
The final wave of \Pythia predictions shows large variations for the $\Delta^{++}(1232)$, $\Sigma^\pm(1385)$, and $\Xi^{0}(1350)$ baryons, despite being specifically matched to the corresponding measurement data.

\begin{figure}[h!]
    \centering
    \includegraphics[width=0.99\linewidth]{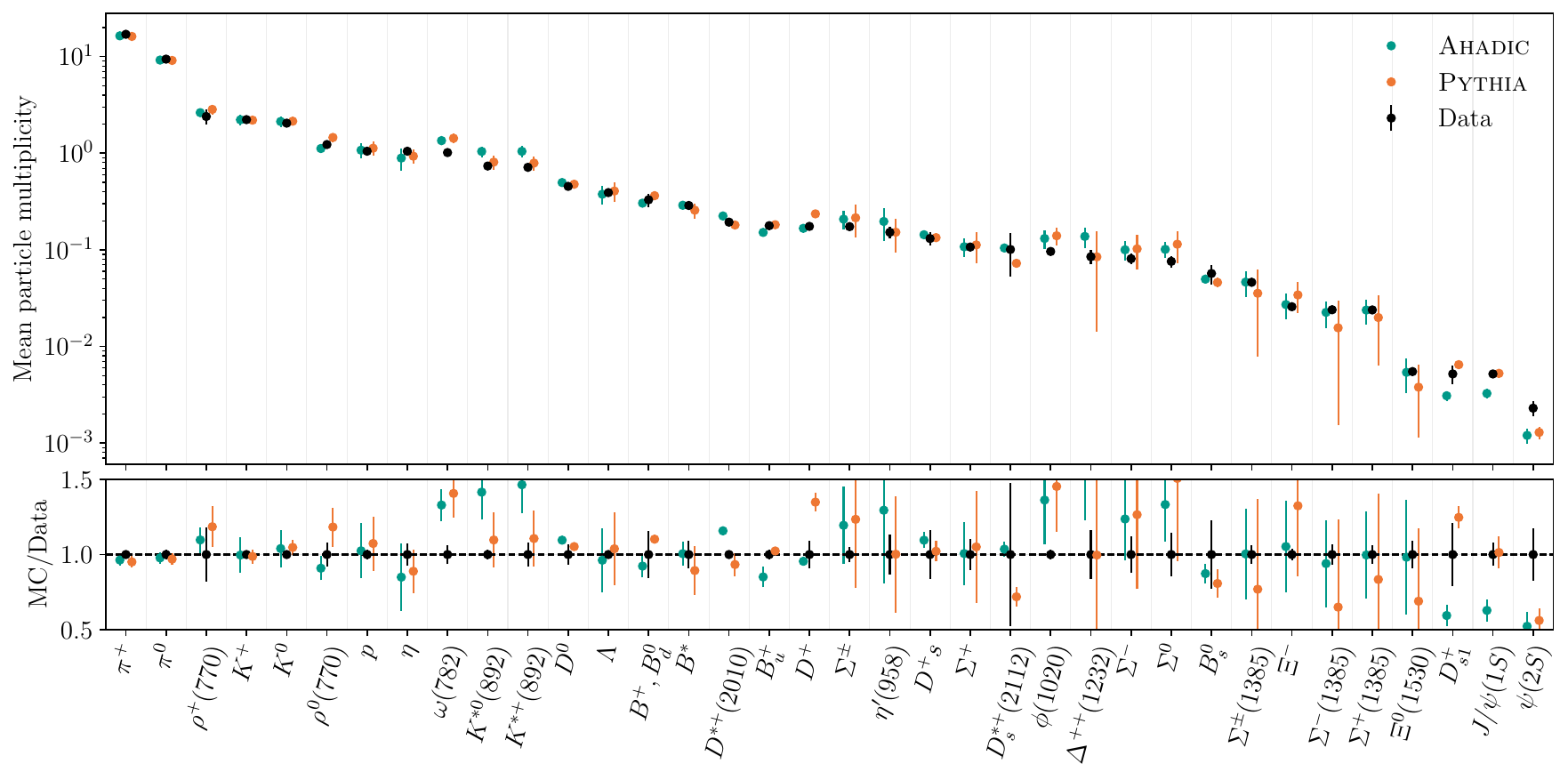}
    \caption{
    Average identified-hadron multiplicities per hadronic $Z$-decay event.
    \Sherpa predictions using \Ahadic (blue) and \Pythia (orange) hadronisation are compared with experimental data compiled by the Particle Data Group~\cite{ParticleDataGroup:2008zun}.
    The bars for the \Ahadic and \Pythia predictions indicate the spread from the final HM wave for each model.
    }
    \label{fig:pdg-hadron-multiplicities}
\end{figure}

In summary, both hadronisation models provide a good overall description of the experimental data.
For inclusive and event-shape observables, the two models yield nearly indistinguishable predictions. The non-perturbative uncertainty envelopes are typically of the order of, or moderately larger than, the experimental uncertainties. However, in particular in the tails of distributions, where also the statistical uncertainty of the MC simulations is non-negligible, 
the parameter uncertainties can significantly exceed the combined measurement and simulation uncertainty. To derive tighter parameter constraints more accurate data and higher statistics in the predictions are needed.  

The most pronounced differences between \Ahadic and \Pythia appear in heavy-flavour fragmentation and in specific hadron-production rates, where the models are sensitive to distinct aspects of their respective flavour and kinematics parametrisations. These residual tensions point to areas where additional data or refined modelling could further constrain the description of non-perturbative effects.

\subsection{Overall $\chi^2$ of the Final Wave Parameter Sets}
\begin{figure}[h!]
    \centering
    \includegraphics[width=0.50\linewidth]{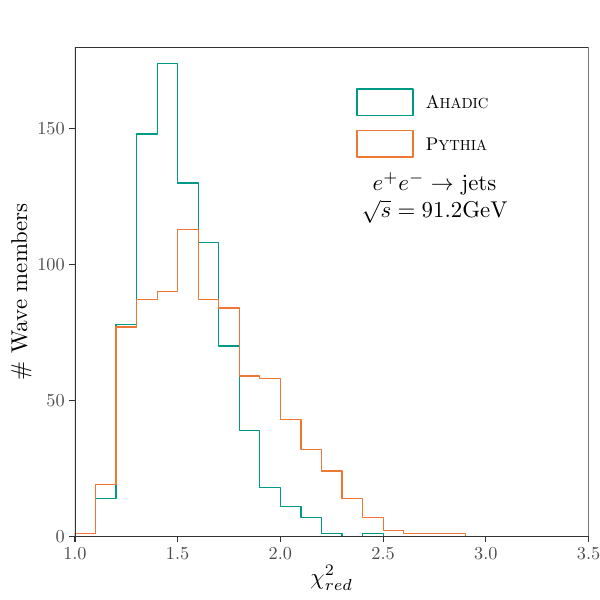}
    \caption{
        Empirical distributions for the reduced $\chi^2$ statistic for the members of the final HM wave for \Ahadic and \Pythia, evaluated for the $432$ observable bins used in the calibration.
    }
    \label{fig:chi2s}
\end{figure}

The members of each wave of HM are chosen to be space-filling samples from the remaining non-implausible parameter space. 
As we have just seen, they can be used to estimate the residual parametric uncertainty of the model. 
In fact, all of them yield a similar performance in describing the data used in the matching procedure and can thus be considered \textit{tunes} of comparable quality.
One must take care to distinguish the nature, and aims, of our measure of implausibility and other metrics of goodness-of-fit that would generate `tunes'. Chief among the differences is that implausibility is never considered as a means of assessing goodness-of-fit; instead, it is only a metric to definitively determine badness-of-fit. The parameter space that remains after successive waves of history matching may, under appropriate distributional assumptions on the model and data, be evaluated via a measure of likelihood: we do so here to facilitate comparison with other established methods of calibration.
We present in Figure~\ref{fig:chi2s} the reduced global $\chi^2$ statistic for the $800$ members of the final HM wave for \Ahadic and \Pythia, respectively: in the notation previously defined, the statistic is calculated as
\begin{equation*}
    \chi^2_{\mathrm{red}} = \frac{1}{N}\sum\limits_{i=1}^N\,\frac{(z_i-f_i(x))^2}{\Var[e_i]+\Var[\epsilon_i(x)]}\,,
\end{equation*}
where here, the quantities $\Var[\epsilon_i]$ and $\Var[e_i]$ must be treated as arising from normal distributions placed on the random variables $\epsilon_i$ and $e_i$. For this comparison, we consider all $N=432$ bins of the $29$ observables used in the calibration. 
Both distributions peak around $\chi^2_\mathrm{red} \approx 1.5$. They are quite narrow, with the \Pythia distribution exhibiting a somewhat more pronounced tail. 
However, none of the wave members exceeds $ \chi^2_{\mathrm{red}}=2.5$ for \Ahadic and $\chi^2_{\mathrm{red}}=3.0$ for \Pythia. 
Accordingly, both models provide a consistent level of agreement with the experimental data across the full ensemble of wave members. 
For comparison, with classical calibration techniques for event-generator tuning, uncertainties are estimated by parameter variations around the \emph{best tune} values that result in $\Delta \chi^2_\mathrm{red}\leq 1$ with respect to the observed (local) minimum~\cite{Buckley:2009bj}.
In the history matching framework, the appropriate induced uncertainty is realised by considering space-filling samples from the entire non-implausible parameter space, i.e., the members of the final wave. 

\subsection{Sampling from the Final Non-Implausible Region}
The above results and considerations of the final non-implausible space provide a host of interesting insights into the model behaviour and the underlying parameters that are compatible with observed reality. 
The trained emulators can be used to generate arbitrarily large numbers of acceptable parameter combinations orders of magnitude faster than would be possible via direct recourse to the simulator, even with appropriate optimisation or Bayesian posterior calibration techniques. 
However, the ultimate goal of many such analyses is not to parametrise the acceptable parameter space. 
For example, one may wish to perform additional analyses on (combinations of) the model output \cite{clark2022tbhiv}; use the model outputs as inputs to a secondary model to produce predictions for future outcomes or as a decision support tool \cite{cavalcante2017history}; or choose future experiments that would be most informative for further restriction of the parameter space. 
If one requires an auxiliary simulator for the task, and in particular if it is computationally expensive, it may be unhelpful to have a large collection of acceptable parameter combinations to propagate through the analysis in order to quantify the uncertainty.

Recall that the non-implausible region is not, strictly speaking, a posterior: each parameter combination chosen is broadly of equal weight in terms of suitability to matching to observational data. 
There is not, therefore, a natural interpretation of a `posterior mode' as would be the case in a fully Bayesian calibration method, for example, and we have no general guarantee that we are able to adequately represent the non-implausible region in terms of reductive summary statistics. 
For instance, the relationship between \texttt{BARYON\_FRACTION} and \texttt{P\_QS\_by\_P\_QQ\_norm} in the \Ahadic non-implausible region shown in Figure~\ref{fig:parameter_space} displays some bimodality, and a naive Gaussian approach would obscure or even misrepresent the relationship between these two variables. 
However, we may still propagate the uncertainty into subsequent analyses in a number of ways, depending on the specifics of the analysis and of the non-implausible region.

If subsequent analyses require a posterior or pseudo-posterior distribution, then we might use the points comprising the final non-implausible space to construct an approximate one. 
If $\hat{\mu}$ is the sample mean of the generated non-implausible points and $\hat{\Sigma}$ the sample variance-covariance matrix, then we construct a proposal distribution
\begin{equation*}
    P(\vect{x}) = \mathcal{N}(\vect{x}; \hat{\mu}, \kappa\hat{\Sigma})
\end{equation*}
where $\kappa > 1$ is some constant chosen to inflate the variance of the non-implausible space to ensure adequate sampling properties. 
Then, defining an approximate likelihood
\begin{equation*}
    L(\vect{x}) = \mathcal{N}(\vect{z}, \Exp_D[g(\vect{x})], \Var_D[g(\vect{x})]+\Var[e]+\Var[\epsilon]),
\end{equation*}
we propose a number of samples $\{\tilde{\vect{x}}\}$ from $P(\vect{x})$, weighting by $w(\vect{x})=L(\vect{x})/P(\vect{x})$, and draw samples $\vect{x}_p$ from $\{\tilde{\vect{x}}\}$ relative to these weights. 
This approach may also be useful if we wish to leverage the reduction of space afforded by HM in another method of calibration such as MCMC, where we may use this approximate posterior as a proposal distribution. 
Care must be taken, however, to ensure that the implicit assumption of normality of the emulator uncertainty is respected (or at least, not flagrantly violated). 
More details of this process can be found in \cite{andrianakis2015bayesian}.

If our later investigation involves future prediction or decision support, it may instead be beneficial to maintain individual parameter combinations generated but subselect so as to have a computationally tractable ensemble of points. 
The nature of the subselection depends on the intended application, but generally we might seek to choose parameter combinations that (as much as possible) represent the full range of output behaviours. 
In simple applications where the outputs are broadly monotonic in the input variables, such a consideration is equivalent to sampling from the convex hull of the non-implausible region.
For more complex behaviours, one can apply Bayesian optimisation with recourse to the emulators to determine those parameter combinations which give rise to the maximal scope of output behaviour~\cite{bordas2020bayesian}. 
Determining in which regime our problem resides would be difficult were we to rely on the computationally expensive simulator runs.
However, as previously shown in Figure~\ref{fig:emexpvar} we may easily explore the expected model behaviour and gain a clear understanding of the model response for any output and combination of parameters.

Finally, we may simply want to identify the single `best' parameter combination, $\vect{x}^*$, akin to classical calibration approaches. 
In this case, we may look directly at relevant measures of goodness-of-fit across the non-implausible space (for example, $\chi^2$ or $L^p$ norm distances) and select appropriately. 
Even in this scenario, we would still wish to parametrise the underlying parametric uncertainty that the history match affords us: since the non-implausible space is representative of the complete space of possible simulator matches to target data we may utilise the ideas behind internal discrepancy to incorporate this information~\cite{goldstein2013assessing}. 
Suppose we have the collection of non-implausible points $\{\vect{x}_{ni}\}=\{\vect{x}_{ni}^{(1)},\dots,\vect{x}_{ni}^{(N)}\}$, and a further simulator $\tilde{f}(\vect{x}, \vect{x^\prime})$ which depends on our matched parameters $\vect{x}$ as well as an additional collection of simulator specific parameters $\vect{x}^\prime$. 
We can perform a set of pilot runs $\{\tilde{f}(\vect{x}_{ni}^{(1)}, \vect{x}^{\prime*}), \dots, f(\vect{x}_{ni}^{(N)}, \vect{x}^{\prime*})\}$ at a fixed point $\vect{x}^{\prime*}$ and use these runs to generate a variance-covariance matrix $\Sigma$ which represents the output variability due to parametric uncertainty in $\vect{x}$. 
This variance matrix can be passed into any subsequent analysis -- for example, a history match on the parameters $\vect{x}^\prime$.

\section{Conclusion}\label{sec:conclude}
In this paper, we have applied, for the first time, the HM method to the calibration of non-perturbative models in the simulation of high-energy collision events with the \Sherpa event-generator framework. 
In particular, we considered the parton-to-hadron transition as modelled by \Sherpa's built-in cluster fragmentation \Ahadic and by the \Pythia Lund-string hadronisation accessed through an interface. 
To constrain the model parameters, we employed a large amount of precision data from the LEP experiments for $e^+e^-\to \text{hadrons}$, comprising $432$ observable bins. 
For our simulations with \Ahadic and \Pythia we treated the perturbative component of the event evolution identically, using NLO QCD matrix elements for the $e^+e^-\to q\bar{q}+\{0,1\}\text{jet}$ processes dressed with \Sherpa's dipole shower. 
Based on very efficient Bayes Linear emulators, HM allowed us to explore the entire parameter space of a given model and successively compress it by exclusion of implausible parameter combinations. 
For the $19$ parameters of the \Ahadic model and the $23$ parameters considered of the \Pythia Lund-string model we found that 3 and 5 consecutive waves of HM, respectively, were sufficient to constrain the model parameters given the uncertainty on the emulators and the data we matched to. 
Analysis of the non-implausible parameter space remaining after the final wave showed that both models are capable of describing the considered LEP data with similar quality, with some minor variations between the two. 
The spread of the observable predictions for the set of parameter points from the final HM wave can be used to estimate the parametric uncertainty of the models. 
By combining the envelopes of the predictions from \Ahadic and \Pythia we can furthermore account for the uncertainty associated with the choice of the physics model for hadronisation. 
An analysis of all two-parameter projections of the \Ahadic and \Pythia parameters revealed strong correlations and multi-modal structures, i.e.\ separated regions in parameter space that provide a similar quality in describing the considered data. 
This highlights a unique capability of HM in contrast to traditional calibration methods that aim to identify a single best-fit parameter point and will typically have problems dealing with multi-modal parameter landscapes.

The HM process allowed us to circumvent the requirement of matching to every output at every wave, and we were able to reduce the number of emulators required at a given wave to around $60$ of the $432$ available.
However, this approach does not completely leverage the structure of the model and its outputs. 
For example, the observations for the $C$-parameter, considered in the context of active variables in Figure~\ref{fig:alephactive}, comprise $50$ bins from what could be considered an implicit density function; one could imagine that these observations could be more efficiently parametrised using hyperparameters of this underlying distribution and so remove the need to consider each in isolation. 
However, such an approach would require a careful consideration of how we might obtain a robust parametrisation of such a distribution in the face of observational uncertainty and carefully propagate any uncertainties through the HM process. 
This remains a future avenue of research. 
Similarly, our consideration of active variables and output subsets at each wave implicitly leveraged our beliefs about the smoothness of and correlations between different bins of the same analysis, but this information might be better employed explicitly via a multivariate emulator and corresponding implausibility measure such as that described in Eq.~\eqref{eq:multivarimp}.
While the burden of specification is higher were we to produce a fully multivariate emulator over each analysis (since we would be required to provide prior specifications for the between-bin covariances), the points proposed from the non-implausible space would inherently respect the trajectory of the predicted histogram outputs, potentially providing good matches to data in fewer waves.

We noted in Section~\ref{sec:SHERPA} and Section~\ref{sec:results} that, in some cases (particularly in those observations in the tail of the histograms) our inability to restrict the parameter space was not only due to emulator uncertainty or observational error, but also due to the inherent stochastic variability of the \Sherpa predictions. 
The use of HM has allowed us to identify the impact of this contribution to the overall uncertainty, and the impact it has on identifying a reduced parameter space of interest.
In this application, the stochastic variability has been inescapable, but in future uses of \Sherpa or where matching to the tails of distributions is paramount we may be motivated to allocate more computational resource, or systematically enhance such rare configurations in the simulation~\cite{Herrmann:2025nnz}, in order to reduce the purely statistical variability.

While there were $432$ outputs of interest here, this does not furnish the entire collection of observational data obtainable from the complete ensemble of LEP analyses. 
Some were omitted from this study because expert judgment deemed those observations to be too unreliable or untrustworthy to draw inference from, or because they are covered by similar measurements already included. 
The results here could be used to validate these assertions, comparing the simulator output in the final non-implausible region to the relevant observational data and potentially highlighting deficiencies or systematic bias in the observation set-up. 
Such considerations may also highlight further differences in behaviour between the \Ahadic and \Pythia models, which have been shown to be broadly consistent across all outputs of interest here.

In this application, our aim has been simply to fully parameterise the non-implausible region, to afford future analyses the opportunity to apply any of the above approaches to propagating uncertainty. 
However, there are other components in particle-level event generators that need to be calibrated using measurement data. 
When considering hadronic collisions for example at the LHC, the fragmentation of hadron remnants and remnant--remnant interactions, called the underlying event, need to be modelled.
We plan to apply HM also for these aspects of collision-event simulation with \Sherpa, paving the way to reliable uncertainty estimations for all non-perturbative aspects of the event evolution. 
Furthermore, to be able to efficiently provide predictions for samples from the corresponding non-implausible parameter space, future work will invest in the development of suitable on-the-fly reweighting algorithms. 
For perturbative uncertainties their in-situ evaluation originating from variations of parameters such as the strong coupling $\alpha_s$ and the renormalisation and factorisation scale in matrix-element and parton-shower simulations has become a widely-used standard technique in modern particle-physics event generators~\cite{Bellm:2016voq,Mrenna:2016sih,Bothmann:2016nao}. 
For non-perturbative models, such reweighting methods are a field of active development~\cite{Bierlich:2023fmh,Assi:2025gog,Butter:2025wxn}. However, they will be instrumental for making uncertainty estimates as derived here practically applicable in phenomenological studies and experimental analyses. 

\section*{Acknowledgements}
This material is based upon work supported by Fermi Forward Discovery Group, LLC under Contract No.\\ 89243024CSC000002 with the U.S. Department of Energy, Office of Science, Office of High Energy Physics. 
A.I.\ is supported by Durham University's Willmore Fellowship.
The work of M.K.\ was supported by the U.S. Department of Energy, Office of Science, Office of Advanced Scientific Computing Research, Scientific Discovery through Advanced Computing (SciDAC-5) program, grant “NeuCol”.  
F.K.\ is grateful for STFC funding under grant agreement ST/P006744/1.
S.S.\ acknowledges financial support from the German Federal Ministry of Research, Technology and Space (BMFTR) through project 05H24MGA.
\FloatBarrier

\printbibliography

\appendix
\section{Parameters and Initial Ranges}\label{app:parameter_selection}
This appendix collects
the parameter ranges used in our calibration. The two tables list, for
each parameter, the corresponding generator setting name and the
scanned initial interval employed in the
study. Table~\ref{tab:params_ahadic} displays the parameters for
\Ahadic, while Table~\ref{tab:params_pythia} shows the corresponding
values for the \Pythia hadronisation model.

\begin{table}[h!]
\centering
\begin{tabular}{p{4cm}|c||p{4cm}|c}
Parameter Name & Range & Parameter Name & Range \\\hline

\texttt{KT\_0}      & [0.5, 1.5] & \texttt{ALPHA\_D}            & [0.0, 4.0] \\
\texttt{ALPHA\_G}   & [0.6, 1.9]  & \texttt{BETA\_D}             & [0.0, 1.0] \\
\texttt{ALPHA\_L}   & [1.0, 4.0]  & \texttt{GAMMA\_D}            & [0.0, 1.0] \\
\texttt{BETA\_L}    & [0.0, 0.4] & \texttt{ALPHA\_H}            & [1.25, 4.0] \\
\texttt{GAMMA\_L}   & [0.1, 0.8]  & \texttt{BETA\_H}             & [0.3, 1.5] \\
\texttt{PT\_MAX}    & [0.5, 3.0]  & \texttt{GAMMA\_H}            & [0.0, 0.4] \\
\texttt{ETA\_MODIFIER}        & [1.5, 4.0] & \texttt{STRANGE\_FRACTION} & [0.2, 0.9] \\
\texttt{ETA\_PRIME\_MODIFIER} & [1.5, 4.0] & \texttt{BARYON\_FRACTION}  & [0.0, 0.7] \\
\texttt{P\_QS\_by\_P\_QQ\_norm} & [0.1, 0.9] & \texttt{P\_SS\_by\_P\_QQ\_norm} & [0.0, 0.1] \\
\texttt{P\_QQ1\_by\_P\_QQ0}     & [0.5, 1.5] & & \\

\end{tabular}
\caption{\Ahadic parameters with the initial maximal ranges considered for the calibration.}
\label{tab:params_ahadic}
\end{table}

\begin{table}[h!]
\centering
\begin{tabular}{l|c||l|c}
Parameter Name & Range & Parameter Name & Range \\\hline

\texttt{StringPT:sigma}            & [0.0, 1.0]  & \texttt{StringZ:aLund}            & [0.0, 2.0] \\
\texttt{StringPT:enhancedFraction} & [0.0, 1.0]  & \texttt{StringZ:bLund}            & [0.0, 2.0] \\
\texttt{StringPT:enhancedWidth}    & [1.0, 10.0] & \texttt{StringZ:aExtraSQuark}     & [-1.0, 2.0] \\
\texttt{StringZ:aExtraDiquark}     & [0.0, 2.0]  & \texttt{StringZ:rFactC}           & [0.0, 2.0] \\
\texttt{StringZ:rFactB}            & [0.0, 2.0]  & \texttt{StringZ:rFactH}           & [0.0, 2.0] \\
\texttt{StringFlav:probStoUD}      & [0.0, 1.0]  & \texttt{StringFlav:probQQtoQ}     & [0.0, 1.0] \\
\texttt{StringFlav:probSQtoQQ}     & [0.0, 1.0]  & \texttt{StringFlav:probQQ1toQQ0}  & [0.0, 1.0] \\
\texttt{StringFlav:mesonUDvector}  & [0.0, 3.0]  & \texttt{StringFlav:mesonSvector}  & [0.0, 3.0] \\
\texttt{StringFlav:mesonCvector}   & [0.0, 3.0]  & \texttt{StringFlav:mesonBvector}  & [0.0, 3.0] \\
\texttt{StringFlav:etaSup}         & [0.0, 1.0]  & \texttt{StringFlav:etaPrimeSup}   & [0.0, 1.0] \\
\texttt{StringFlav:popcornRate}    & [0.0, 2.0]  & \texttt{StringFlav:popcornSpair}  & [0.0, 1.0] \\
                                    &             & \texttt{StringFlav:popcornSmeson} & [0.0, 1.0] \\

\end{tabular}
\caption{\Pythia 8 hadronisation parameters together with the initial maximal ranges considered for the calibration.}
\label{tab:params_pythia}
\end{table}

\section{Reference Data Selection}\label{app:observable_selection}
In this appendix we give a brief overview of the experimental data used to calibrate our models.
All analyses were used via \Rivet. Table~\ref{tab:analysis} lists the corresponding \Rivet analysis tag, the associated publication, and the observables employed for the calibration.

\begin{table}[h!]
    \centering
    \begin{tabular}{l|c|l}
      \Rivet tag &  Ref. & Observables \\\hline
      \texttt{ALEPH\_2001\_I558327 }  & \cite{ALEPH:2001pfo} & $b$-quark fragmentation \\
      \texttt{ALEPH\_2004\_I636645 }  & \cite{ALEPH:2003obs} & Jet rates, event-shape variables, \\
      & & Inclusive charged-particle spectra\\
      \texttt{ALEPH\_1991\_I319520 }       & \cite{ALEPH:1991ldi}  & Charged-particle multiplicity \\
      \texttt{ALEPH\_1999\_I507422 }       & \cite{ALEPH:1999syy}  & Scaled-energy distribution of $D^*$ \\
      \texttt{DELPHI\_1996\_I424112 }      & \cite{DELPHI:1996sen} & Event-shape variables, differential jet rates\\
      \texttt{OPAL\_1994\_I372772 }        & \cite{OPAL:1994zan}   & Inclusive production rates of $\pi^\pm$, $K^\pm$, $p$, $\bar{p}$ \\
      \texttt{OPAL\_1998\_I472637 }        & \cite{OPAL:1998arz}   & Scaled-momentum distributions and \\
      & & total charged multiplicities in flavour-tagged events\\
      \texttt{OPAL\_2000\_I529898 }        & \cite{OPAL:2000dkf}   & Multiplicities of $\pi^0$, $\eta$ and $K^0$  \\
       & & and of charged particles in quark and gluon jets\\
      \texttt{OPAL\_2003\_I599181 }        & \cite{OPAL:2002plk}   & $b$-quark fragmentation \\
      \texttt{SLD\_2004\_I630327 }         & \cite{SLD:2003ogn}    & Differential production rates of $\pi^\pm$, $K^\pm$ and $p,\bar{p}$ \\   
      \texttt{PDG\_HADRON\_MULTIPLICITIES} & \cite{ParticleDataGroup:2008zun} & Various meson and baryon multiplicities \\
    \end{tabular}
    \caption{Experimental data used in the calibration.}
    \label{tab:analysis}
\end{table}

\section{Auxiliary Results}\label{app:additional_figures}
\begin{figure}[h!]
  \centering
  \includegraphics[width=\textwidth]{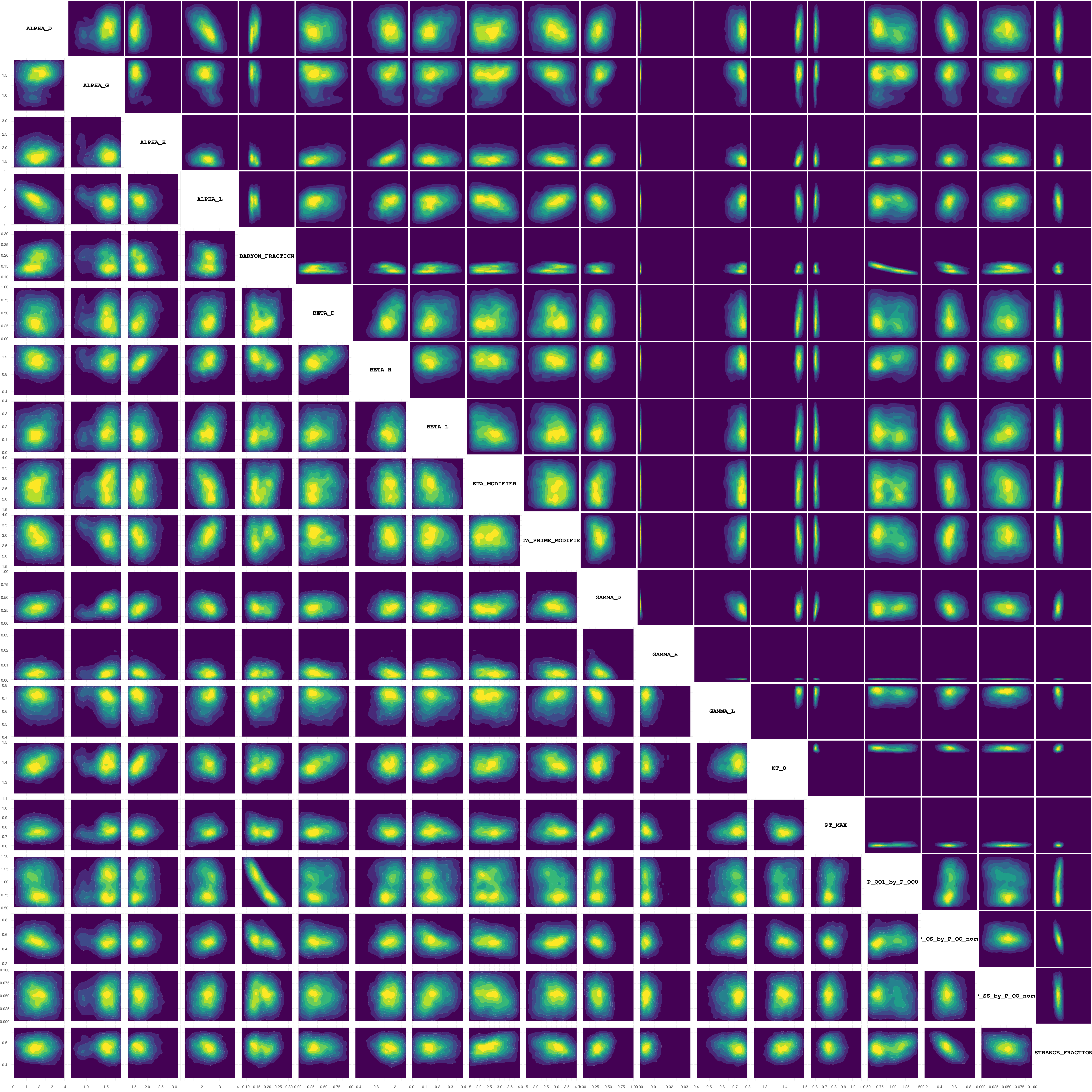}
  \caption{Two-dimensional projections of the non-implausible \Ahadic parameter space at the final HM wave, showing all pairwise correlations. Plots in the upper triangle show correlations relative to the original parameter volume, while those in the lower triangle are focused on the final plausible region. The densities are estimated from a  sample of $10^4$ non-implausible points.
  }
  \label{fig:2d_ahadic_correlations}
\end{figure}

\begin{figure}[h!]
  \centering  
  \includegraphics[width=\textwidth]{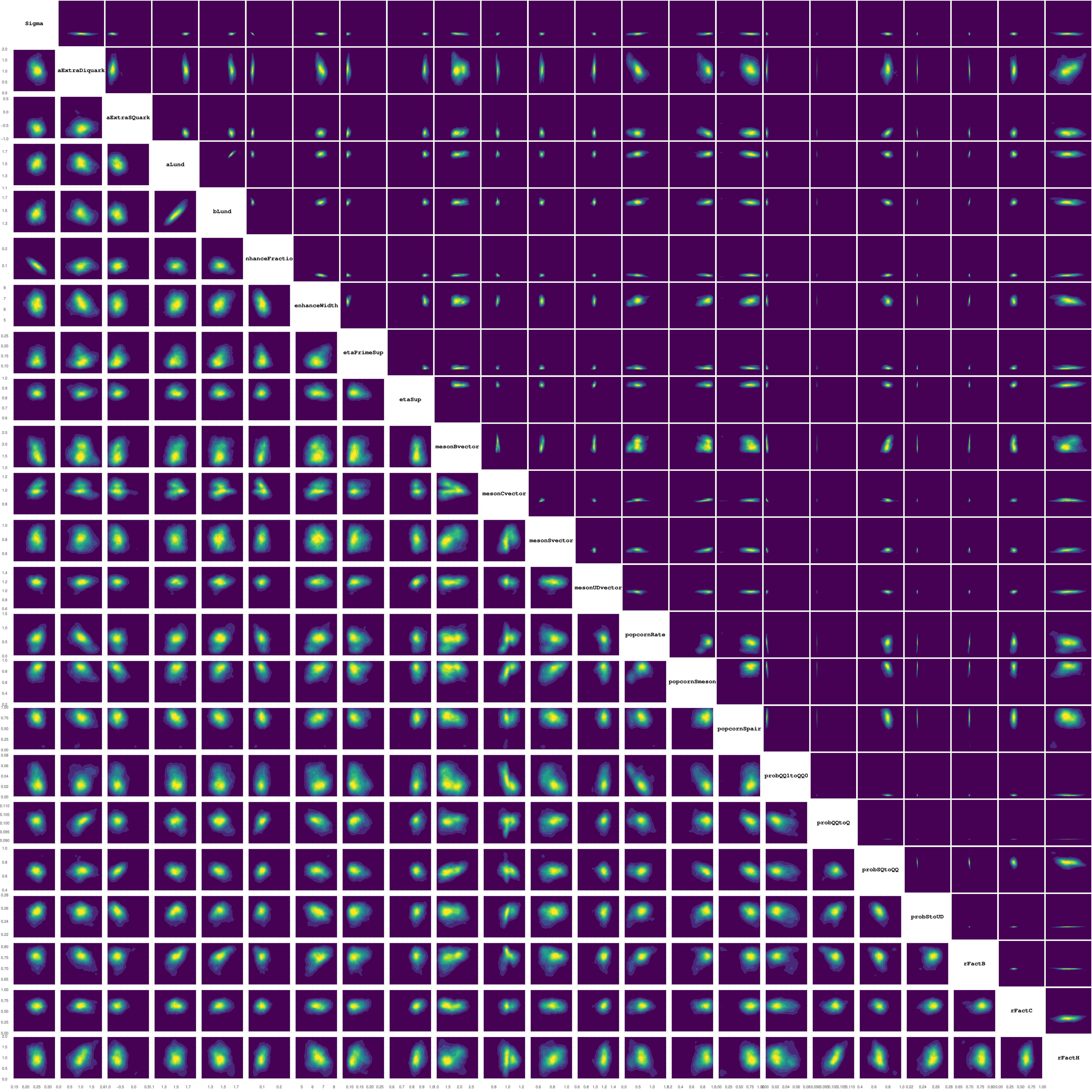}
  \caption{Two-dimensional projections of the non-implausible \Pythia parameter space at the final HM wave, showing all pairwise correlations. Plots in the upper triangle show correlations relative to the original parameter volume, while those in the lower triangle are focused on the final plausible region. The densities are estimated from a  sample of $10^4$ non-implausible points.
  }
  \label{fig:2d_pythia_correlations}
\end{figure}

\begin{figure}
    \centering
    \includegraphics[width=0.32\linewidth]{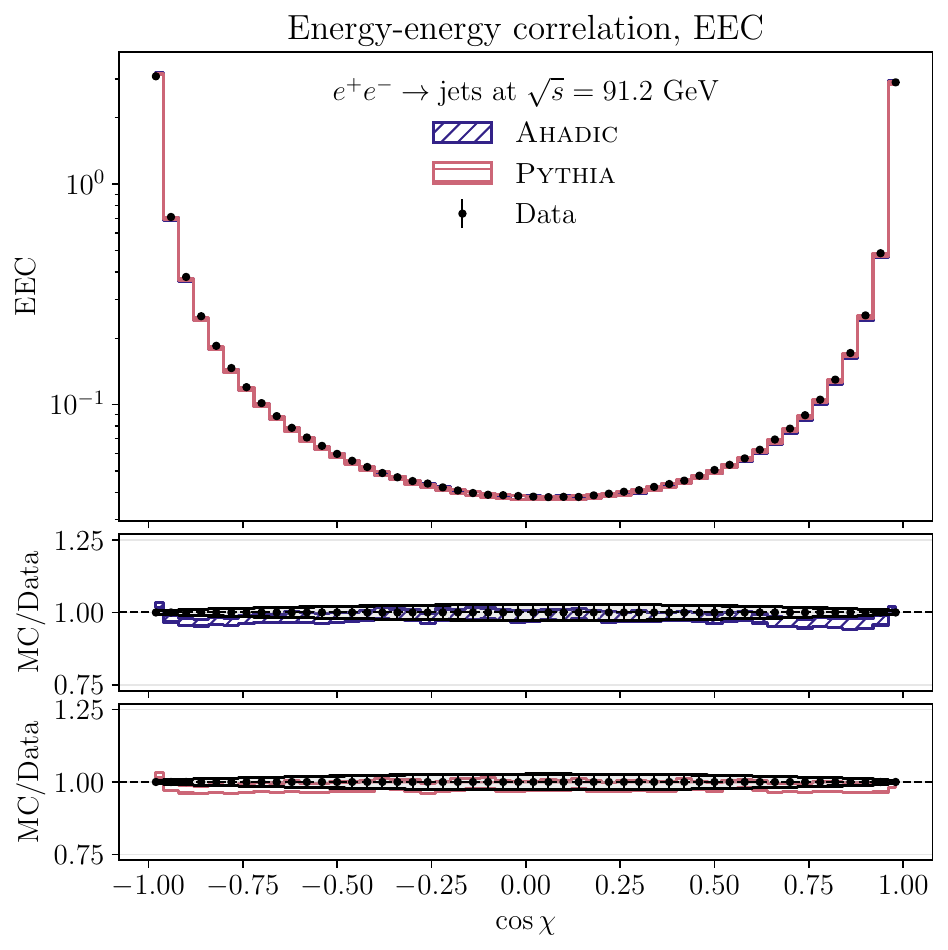}
    \includegraphics[width=0.32\linewidth]{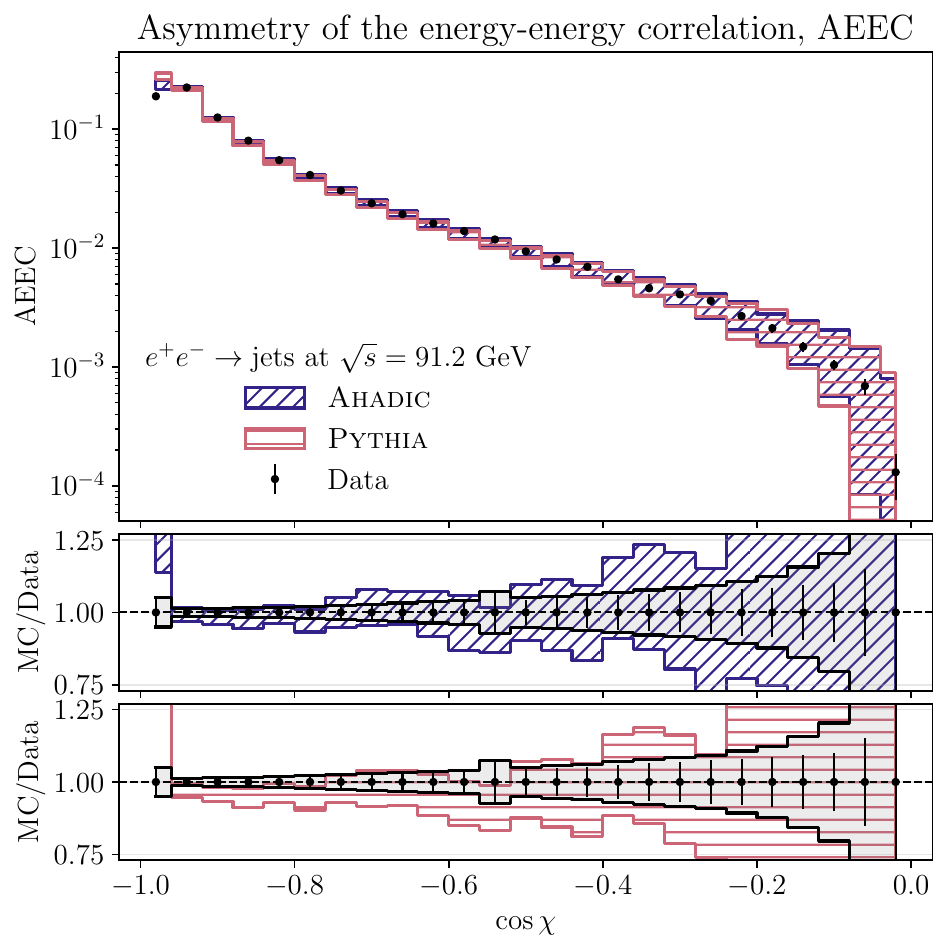}
    \caption{
        \Sherpa results comparing \Ahadic and \Pythia
        hadronisation for the Energy--Energy Correlation (EEC) and its asymmetry (AEEC).
        The experimental data is taken from~\cite{DELPHI:1996sen}. 
        The coloured shaded bands represent the envelopes spanned by all parameter sets in the final HM wave for each model, while the black band indicates the combined error of the experimental data (black markers) and the average statistical uncertainty of the simulator predictions. 
    }
    \label{fig:energycorrelators}
\end{figure}

\begin{figure}
    \centering
    \includegraphics[width=0.32\linewidth]{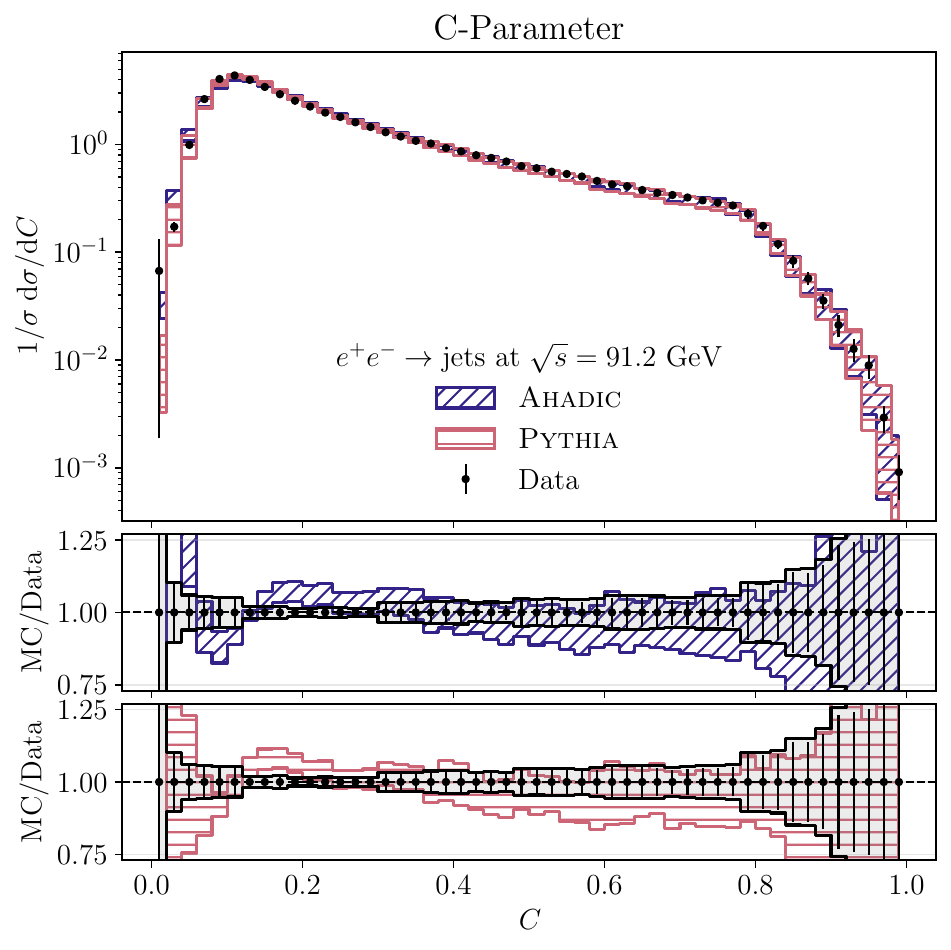}
    \includegraphics[width=0.32\linewidth]{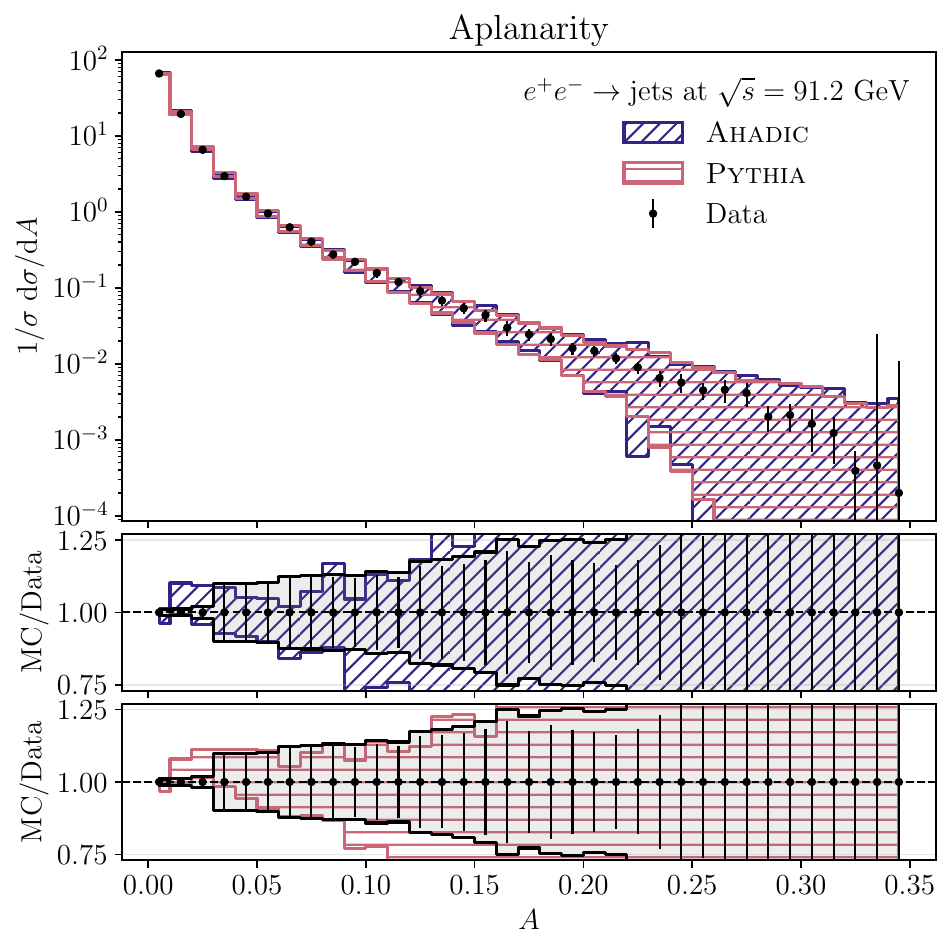}
    \includegraphics[width=0.32\linewidth]{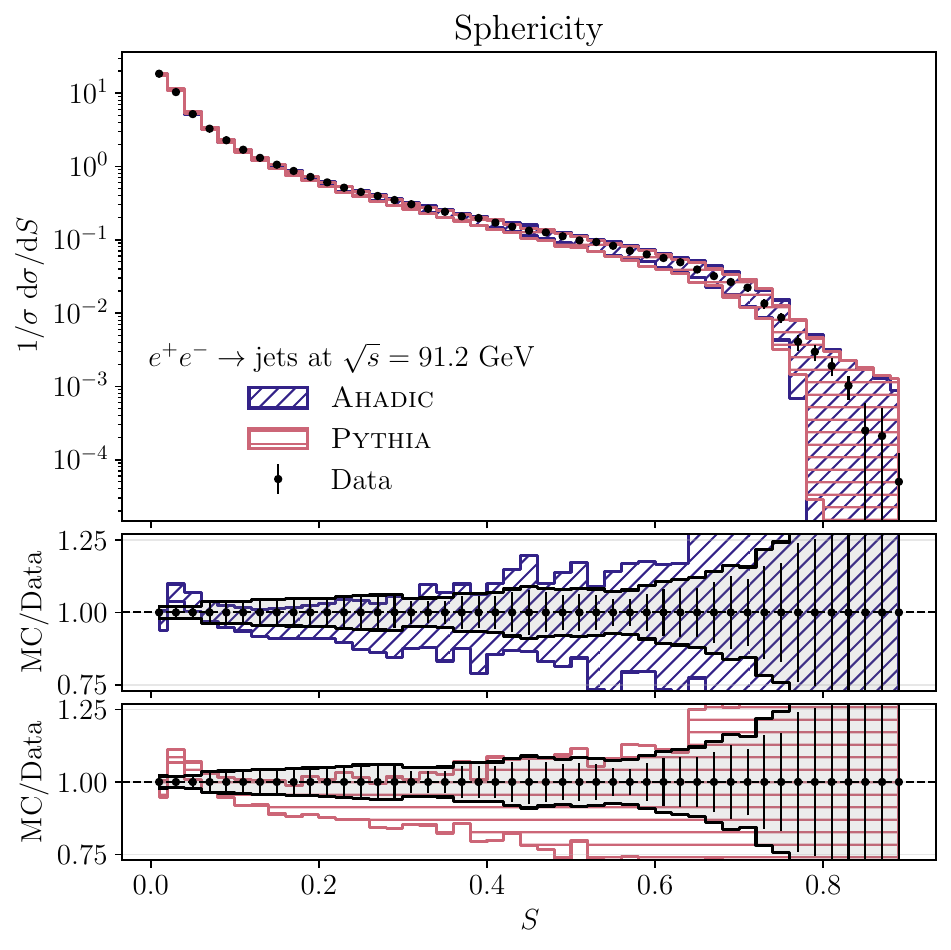}
    \caption{Distributions of several event-shape observables measured
      in $e^+e^-$ annihilation to hadrons. Left to right: 
      $C$-parameter, aplanarity and sphericity, measured by ALEPH~\cite{ALEPH:2003obs}. The coloured shaded bands represent the envelopes spanned by all parameter sets in the final HM wave for each model, while the black band indicates the combined error of the experimental data (black markers) and the average statistical uncertainty of the simulator predictions. 
    }
    \label{fig:event_shapes}
\end{figure}

\begin{figure}
    \centering
    \includegraphics[width=0.32\linewidth]{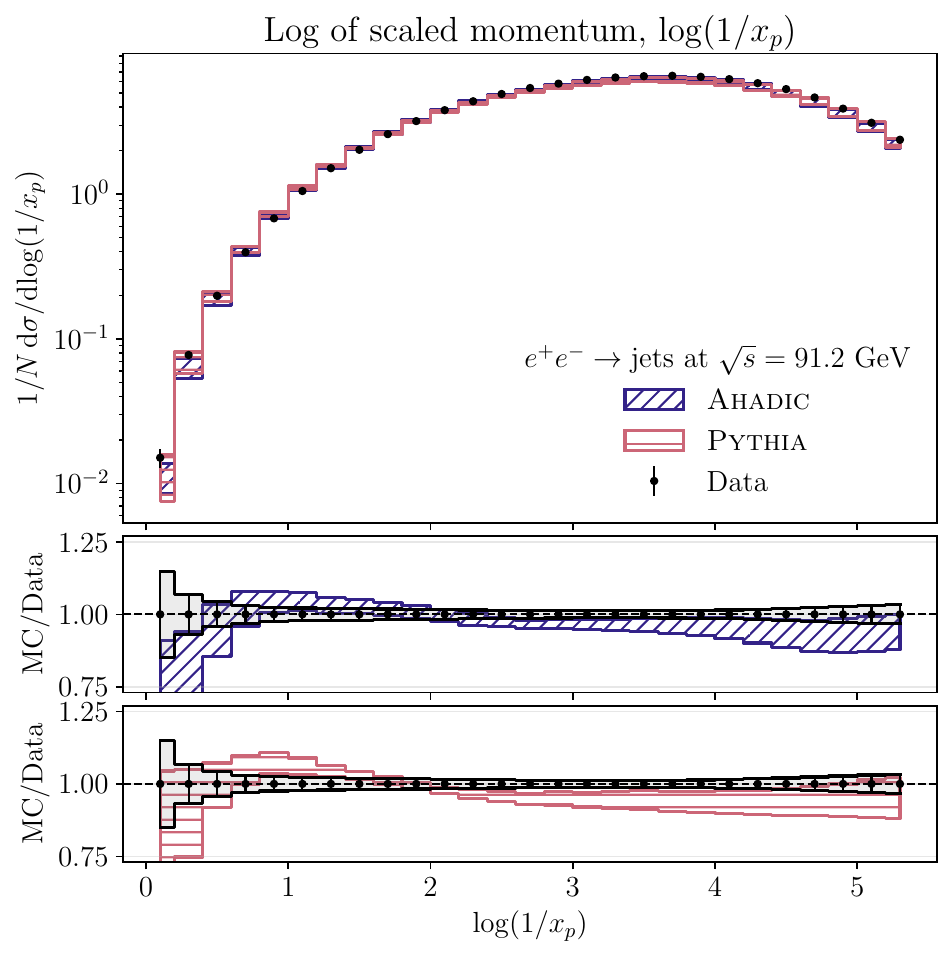}
    \includegraphics[width=0.32\linewidth]{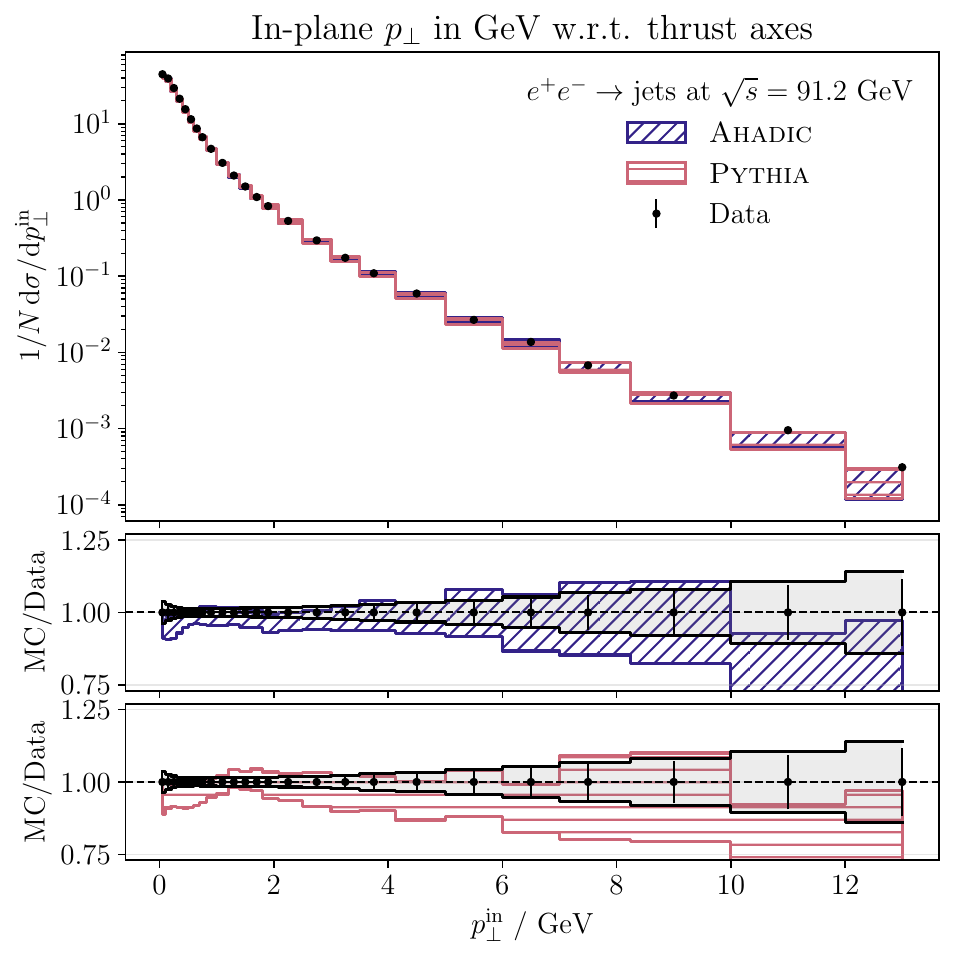}
    \includegraphics[width=0.32\linewidth]{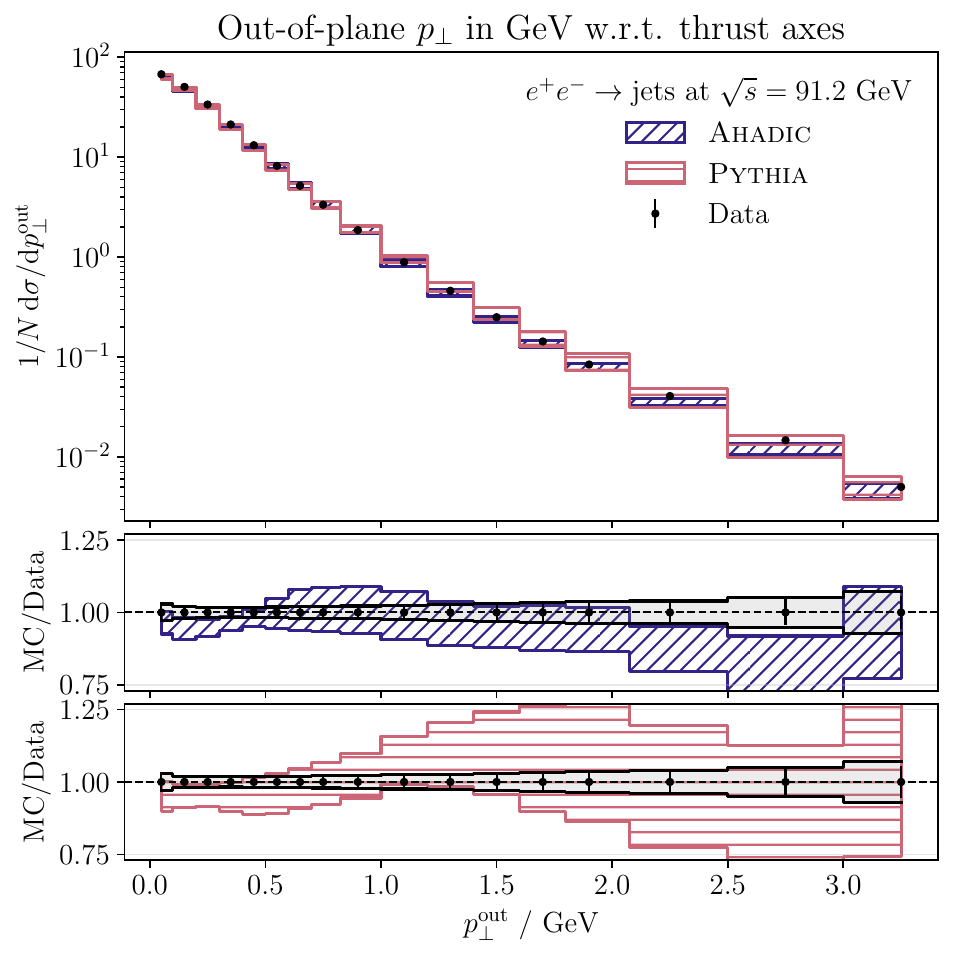}
    \includegraphics[width=0.32\linewidth]{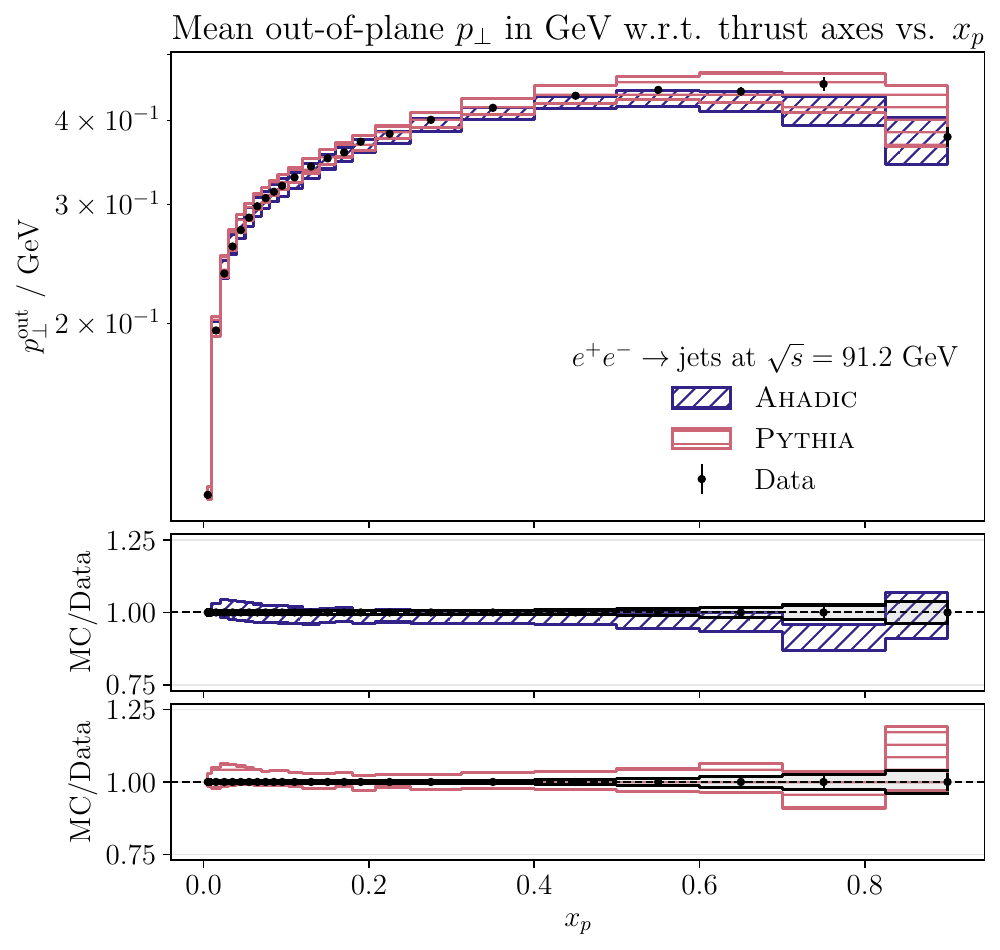}
    \includegraphics[width=0.32\linewidth]{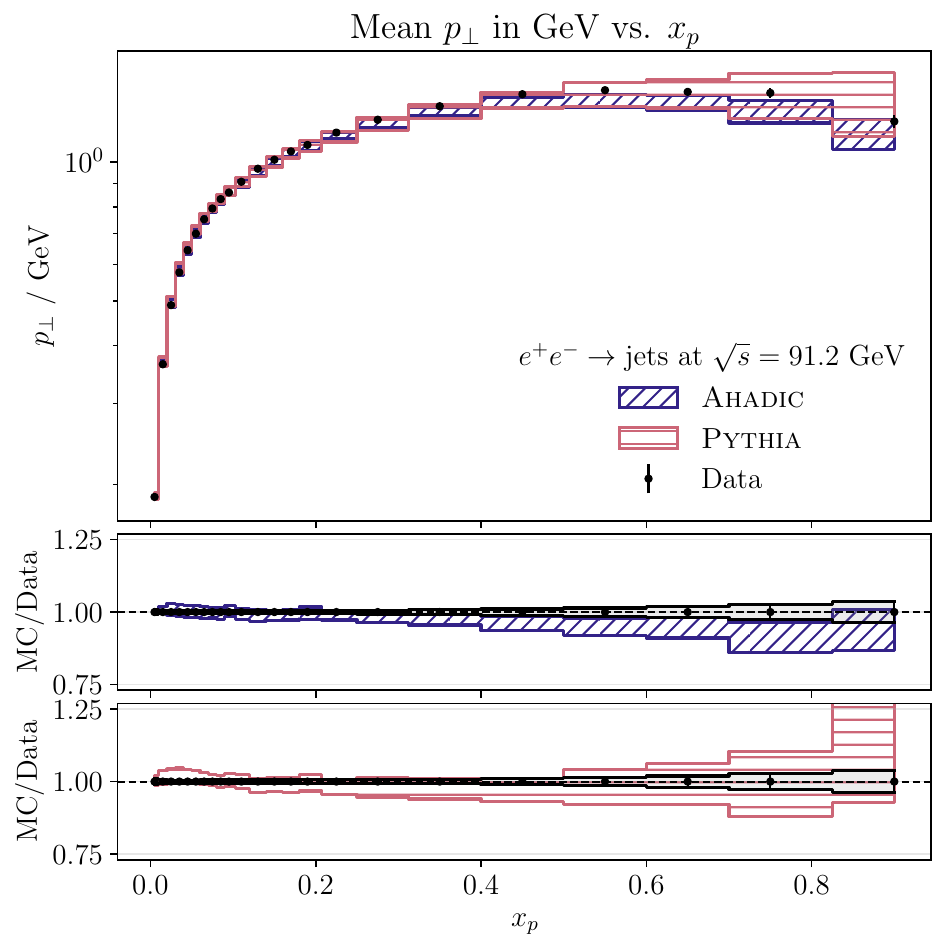}
    \includegraphics[width=0.32\linewidth]{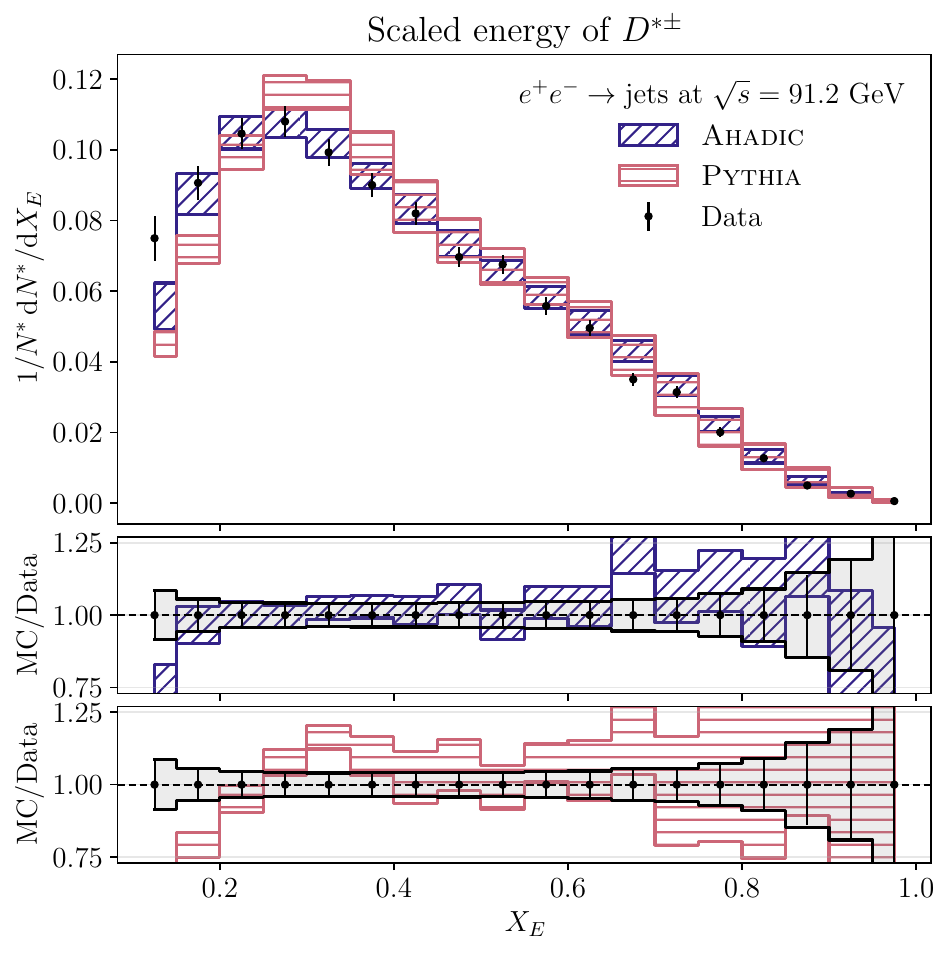}
    \includegraphics[width=0.32\linewidth]{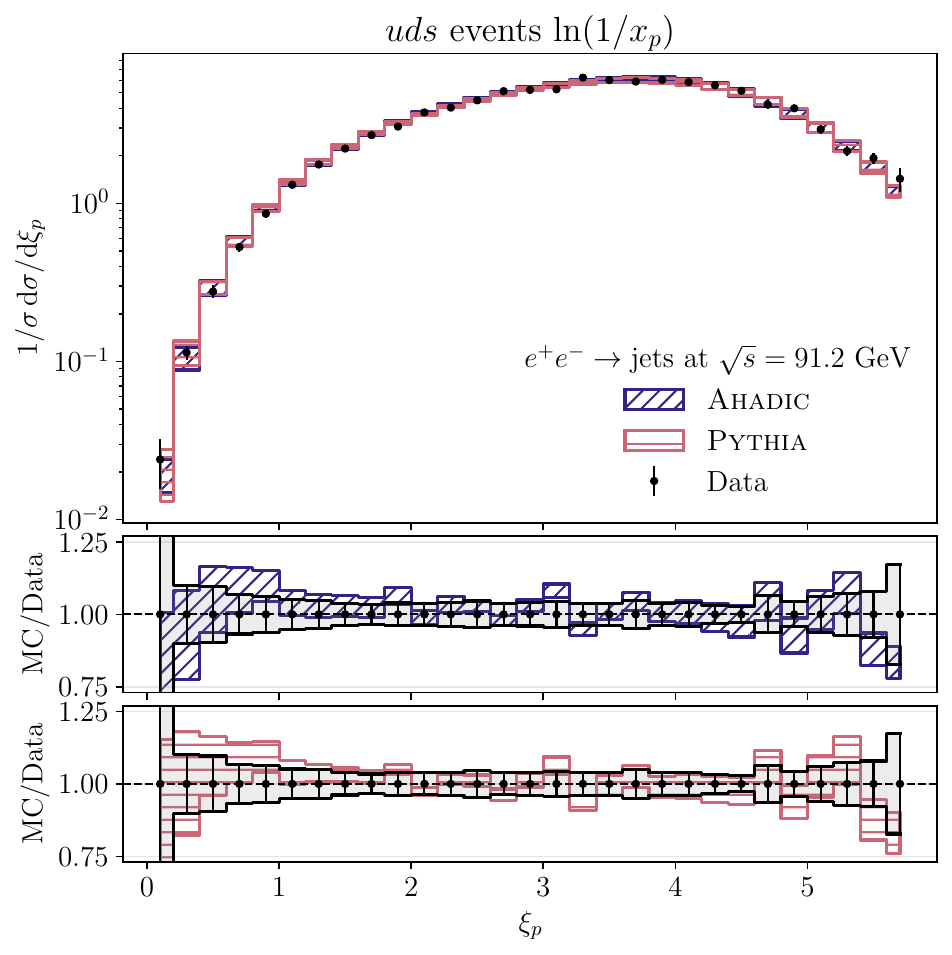}
    \includegraphics[width=0.32\linewidth]{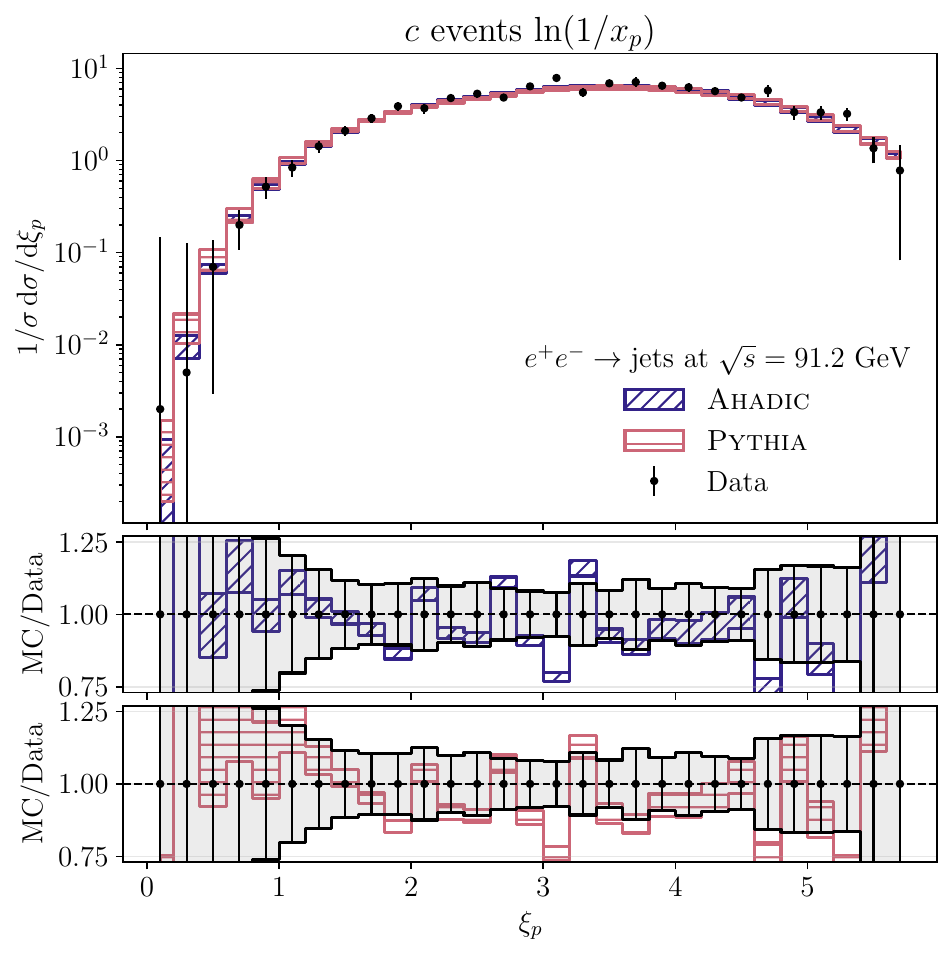}
    \includegraphics[width=0.32\linewidth]{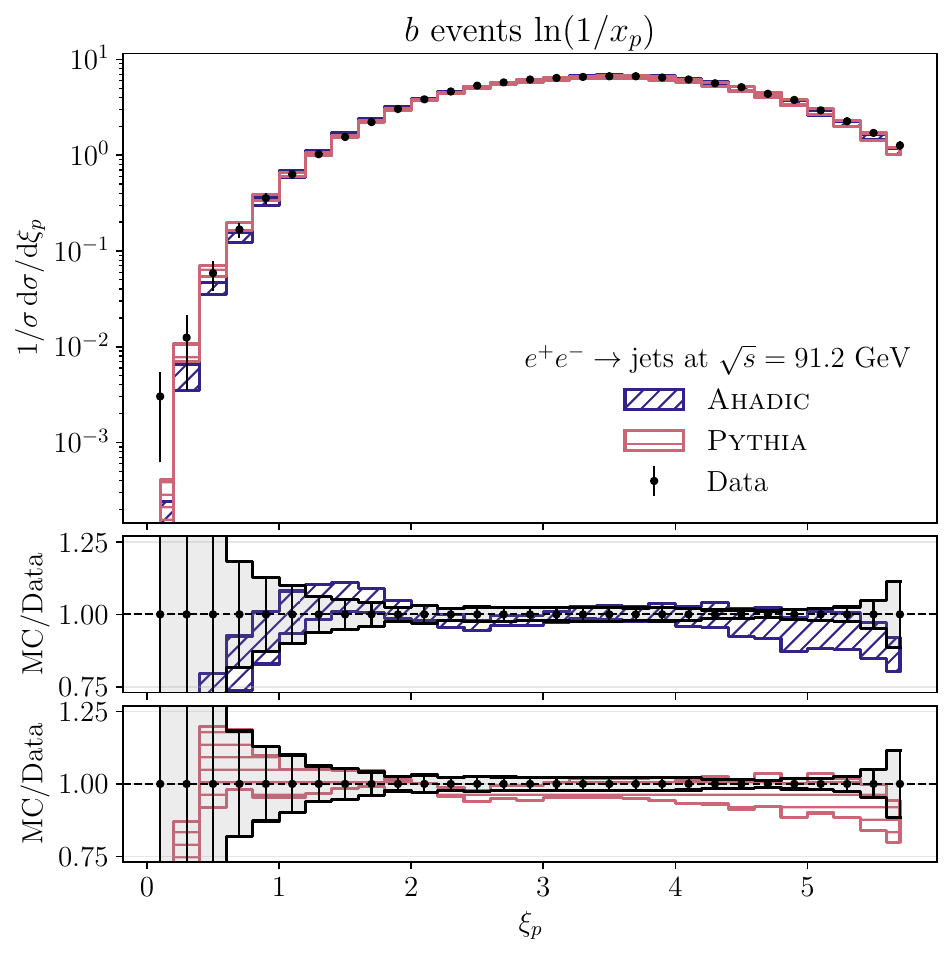}
 
    \caption{\Sherpa results comparing \Ahadic and \Pythia
      hadronisation for longitudinal momentum fractions, transverse
      momenta in- and out-of-plane w.r.t. the thrust axis to
      experimental data~\cite{DELPHI:1996sen}.  We also show the 
      scaled-energy distribution of $D^*$ mesons with 
      experimental data from~\cite{ALEPH:1999syy} and the scaled
      momentum distributions split by flavour, i.e. separately for
      $uds$, $c$ and $b$ quarks with experimental data taken
      from~\cite{OPAL:1998arz}. The coloured shaded bands represent the envelopes spanned by all parameter sets in the final HM wave for each model, while the black band indicates the combined error of the experimental data (black markers) and the average statistical uncertainty of the simulator predictions. 
      }
    \label{fig:particle_spectra}
\end{figure}

In this appendix we collate supplementary material concerning the application of HM for the \Ahadic
and \Pythia hadronisation models. In Figure~\ref{fig:2d_ahadic_correlations} we show two-parameter projections of the final non-implausible space for the case of \Ahadic. The corresponding results for \Pythia can be found in Figure~\ref{fig:2d_pythia_correlations}. Finally, Figures~\ref{fig:energycorrelators},~\ref{fig:event_shapes}, and \ref{fig:particle_spectra} present additional comparisons of \Ahadic and \Pythia predictions for parameter sets from the final HM wave with 
experimental data. 

\end{document}

%% file: macros/macros.tex
\usepackage{amsmath}
\usepackage{amssymb}
\usepackage{array}
\usepackage{calc}
\usepackage{longtable}
\usepackage{multirow,booktabs}
\usepackage{relsize}
\usepackage{pstricks}
\usepackage{graphicx}
\usepackage{xspace}
\usepackage{units}
\usepackage{afterpage}
\usepackage{placeins}

\numberwithin{equation}{section}

\usepackage[format=hang,labelfont=bf,hypcap=true]{caption}
\usepackage{subcaption}
\usepackage{bm}
\let\spreprint\empty
\newcommand{\preprint}[1]{\def\spreprint{\protect#1}}
\let\sinstitute\empty
\newcommand{\institute}[1]{\def\sinstitute{\protect#1}}
\makeatletter
\renewcommand{\maketitle}{\begingroup
  \null\thispagestyle{empty}%
    \ifx\spreprint\empty
      \vskip 5ex
    \else
      \flushright\large\spreprint\vskip 2ex
    \fi
    \vskip 5ex
    \flushleft
      {\bfseries\huge\@title}\vskip 6ex
      \@author\vskip 2ex
      \ifx\sinstitute\empty
      \else
        {\small\sinstitute}
      \fi
    \vskip 5ex
  \endgroup
}
\makeatother

\numberwithin{equation}{section}

\input{macros/macros_physics}

\input{macros/macros_maths}

\input{macros/macros_refs}

\input{macros/macros_collabs}

\input{macros/macros_codes}

%% file: macros/macros_physics.tex
\newcommand{\UGeV}{\ensuremath{\mathrm{GeV}}\xspace}

\newcommand{\sla}[1]{\ensuremath{{#1\kern-0.45em/}}}

%% file: macros/macros_maths.tex
\newcommand{\bea}{\begin{align}}
\newcommand{\eea}{\end{align}}
\newcommand{\beq}{\begin{equation}}
\newcommand{\eeq}{\end{equation}}

\newcommand{\bs}{\begin{split}}
\newcommand{\es}{\end{split}}

\newcommand{\bi}{\begin{itemize}}
\newcommand{\ei}{\end{itemize}}
\newcommand{\bc}{\begin{center}}
\newcommand{\ec}{\end{center}}
\newcommand{\bac}{\begin{array}{c}}
\newcommand{\bacc}{\begin{array}{cc}}
\newcommand{\ea}{\end{array}}

\def\spa#1.#2{\langle#1\,#2\rangle}
\def\spb#1.#2{[#1\,#2]}


%% file: macros/macros_collabs.tex
\newcommand\LHC{L\protect\scalebox{0.8}{HC}\xspace}

%% file: macros/macros_codes.tex
\newcommand{\MEPSatNLO}{M\scalebox{0.8}{E}P\scalebox{0.8}{S}@N\protect\scalebox{0.8}{LO}\xspace}

\newcommand{\Herwig}{H\protect\scalebox{0.8}{ERWIG}\xspace}

\newcommand{\Pythia}{P\protect\scalebox{0.8}{YTHIA}\xspace}

\newcommand{\Sherpa}{S\protect\scalebox{0.8}{HERPA}\xspace}

\newcommand{\Ahadic}{A\protect\scalebox{0.8}{HADIC}\xspace}

\newcommand{\Rivet}{R\protect\scalebox{0.8}{IVET}\xspace}